\documentclass{aa}  

\usepackage{txfonts}
\usepackage{multirow}
\usepackage{subfig,graphicx}

\usepackage{natbib,twoopt}
\usepackage[breaklinks=true]{hyperref} 
\bibpunct{(}{)}{;}{a}{}{,}             
\makeatletter
  \newcommandtwoopt{\citeads}[3][][]{\href{http://adsabs.harvard.edu/abs/#3}%
    {\def\hyper@linkstart##1##2{}%
     \let\hyper@linkend\@empty\citealp[#1][#2]{#3}}}
  \newcommandtwoopt{\citepads}[3][][]{\href{http://adsabs.harvard.edu/abs/#3}%
    {\def\hyper@linkstart##1##2{}%
     \let\hyper@linkend\@empty\citep[#1][#2]{#3}}}
  \newcommandtwoopt{\citetads}[3][][]{\href{http://adsabs.harvard.edu/abs/#3}%
    {\def\hyper@linkstart##1##2{}%
     \let\hyper@linkend\@empty\citet[#1][#2]{#3}}}
  \newcommandtwoopt{\citeyearads}[3][][]%
    {\href{http://adsabs.harvard.edu/abs/#3}
    {\def\hyper@linkstart##1##2{}%
     \let\hyper@linkend\@empty\citeyear[#1][#2]{#3}}}
\makeatother

\begin{document}

   \title{Identifying type II quasars at intermediate redshift with few-shot learning photometric classification}
    \titlerunning{Identifying QSO2 at 1 $\leq$z $\leq$ 2 with few-shot learning photometric classification}

   \author{P.~A.~C. Cunha\inst{1,2} \fnmsep\thanks{Corresponding author, \email{pedro.cunha@astro.up.pt}}
          \and
          A. Humphrey\inst{2,3}
          \and
          J. Brinchmann\inst{1,2}
          \and
          S.~G. Morais\inst{1,2}
         \and
          R. Carvajal\inst{4,5}
         \and
          J.~M. Gomes\inst{2}
          \and
          I. Matute\inst{4,5}
         \and
          A. Paulino-Afonso\inst{2}
          }

    \institute{Departamento de Física e Astronomia, Faculdade de Ciências, Universidade do Porto, Rua do Campo Alegre 687, PT4169-007 Porto, Portugal
         \and
             Instituto de Astrofísica e Ciências do Espaço, Universidade do Porto, CAUP, Rua das Estrelas, PT4150-762 Porto, Portugal
        \and 
            DTx – Digital Transformation CoLab, Building 1, Azurém Campus, University of Minho, PT4800-058 Guimarães, Portugal
        \and
            Departamento de Física, Faculdade de Ciências, Universidade de Lisboa, Edifício C8, Campo Grande, PT1749-016 Lisboa, Portugal
        \and
            Instituto de Astrofísica e Ciências do Espaço, Universidade de Lisboa, OAL, Tapada da Ajuda, PT1349-018 Lisboa, Portugal
             }

   \date{Received ; accepted }

   \abstract
   {A sub-population of AGNs where the central engine is obscured are known as type II quasars (QSO2s). These luminous AGNs have a thick and dusty torus that obscures the accretion disc from our line of sight. Thus, their special orientation allows for detailed studies of the AGN-host co-evolution. Increasing the sample size of QSO2 sources in critical redshift ranges is crucial for understanding the interplay of AGN feedback, the AGN-host relationship, and the evolution of active galaxies.}
   { We aim to identify QSO2 candidates in the `redshift desert' using optical and infrared photometry. At this intermediate redshift range (i.e. $1 \leq z \leq 2$), most of the prominent optical emission lines in QSO2 sources (e.g. CIV$\lambda1549$; [OIII]$\lambda\lambda 4959, 5008$) fall either outside the wavelength range of the SDSS optical spectra or in particularly noisy wavelength ranges, making QSO2 identification challenging. Therefore, we adopted a semi-supervised machine learning approach to select candidates in the SDSS galaxy sample.}
   {Recent applications of machine learning in astronomy focus on problems involving large data sets, with small data sets often being overlooked. We developed a `few-shot' learning approach for the identification and classification of rare-object classes using limited training data (200 sources). The new \texttt{AMELIA} pipeline uses a transfer-learning based approach with decision trees, distance-based, and deep learning methods to build a classifier capable of identifying rare objects on the basis of an observational training data set.}
   {We validated the performance of \texttt{AMELIA} by addressing the problem of identifying QSO2s at $1 \leq z \leq 2$ using SDSS and WISE photometry, obtaining an F1-score above 0.8 in a supervised approach. We then used \texttt{AMELIA} to select new QSO2 candidates in the `redshift desert' and examined the nature of the candidates using SDSS spectra, when available. In particular, we identified a sub-population of [NeV]$\lambda3426$ emitters at $z\sim1.1$, which are highly likely to contain obscured AGNs. We used X-ray and radio cross-matching to validate our classification and investigated the performance of photometric criteria from the literature showing that our candidates have an inherent dusty nature. Finally, we derived physical properties for our QSO2 sample using photoionisation models and verified the AGN classification using an SED fitting.}
   {Our results demonstrate the potential of few-shot learning applied to small data sets of rare objects, in particular QSO2s, and confirms that optical-IR information can be further explored to search for obscured AGNs. We present a new sample of candidates to be further studied and validated using multi-wavelength observations.}

   \keywords{obscured quasars -- galaxies -- photometry -- spectroscopy -- machine learning}

   \maketitle
%
\section{Introduction}

The study of the Universe around us is extremely dependent on our understanding of galaxies since various key processes, such as star formation, accretion, and feedback are present on galaxy scales \citep{1998A&A...331L...1S,2000ApJ...539L...9F,2000ApJ...539L..13G,2012ARA&A..50..455F, 2006ApJS..163....1H}. Active galactic nuclei (AGNs) -- supermassive black holes (SMBHs) powered by gas accretion \citep{1969Natur.223..690L} -- are particularly relevant in this context due to their often significant impact on the properties of their host galaxies \citep[e.g.][]{1997ApJ...487L.105C,2003MNRAS.346.1055K,2006MNRAS.370..645B,2008ApJ...673..715C,2008ApJ...675.1025S,2013MNRAS.429.1827P,2018A&A...612A..31P,10.1093/mnras/staa045,10.1093/mnras/staa2222,2021IAUS..359..203H,2022MNRAS.510.4485R,2023MNRAS.tmp...14E}.

From the panoply of currently known AGN classes, obscured AGNs -- or type II quasars (QSO2s) -- are particularly useful for studying AGN hosts. According to the unified AGN model \citep{1993ARA&A..31..473A, 1995PASP..107..803U,2017NatAs...1..679R}, the luminous central engine of QSO2s is hidden from view by what is typically argued to be a fortuitously orientated dusty torus \citep[e.g.][]{1988ApJ...328..161K}. This favourable geometry allows the host galaxy to be studied in greater detail, compared to the case of unobscured quasars (QSO1s), where the active nucleus often outshines the host galaxy. In some obscured AGNs, infrared and sub-millimetre observations have hinted at a different scenario in which the obscuration of the central engine is caused by a two-component dusty structure with equatorial and polar components, with radiation pressure playing an important rule \citep[][and references therein]{2019ApJ...884..171H, 2023MNRAS.519.3237S}. Thus, the selection and study of obscured AGNs are essential for understanding the AGN population, the co-evolution of AGNs with their host galaxies, and other open issues, such as the radiative efficiency of SMBHs \citep{2018ARA&A..56..625H}.

Since their discovery in significant numbers in the Sloan Digital Sky Survey \citep[SDSS;][]{2004AJ....128.1002Z}, QSO2s have been studied in various wavelengths and with various observational techniques. They typically have an optical emission line spectrum consisting of narrow lines from low-ionisation species (e.g. [OII]$\lambda\lambda3727,3730$) and high-ionisation species (e.g. [OIII]$\lambda\lambda4959,5008$) in addition to recombination lines such as H$\alpha$. The luminosities of their narrow forbidden lines are comparable to those found in QSO1s \citep[e.g][]{2008AJ....136.2373R}. Moreover, IR observations have played a crucial role in the selection of active galaxies, including QSO2s, across different redshifts \citep[e.g.][]{2005ApJ...631..163S, 2006MNRAS.370.1479M,2008AJ....136.1607Z}. X-ray surveys have also detected this type of source, with L$_X \geq 10^{44}$ erg s$^{-1}$ \citep[e.g.][]{2003AJ....126..632B, 2004ApJS..155..271S, 2004AJ....128.1002Z, 2005A&A...431...87S, 2006ApJ...637..147P, 2006A&A...451..859S,2007A&A...467...73T}, revealing a wide range of X-ray luminosities and HI column densities. For example, in \citet{2006ApJ...637..147P}, five of the eight studied QSO2s showed column densities consistent with Compton-thick sources (i.e. N$_H \geq 10^{22}$ cm$^{-2}$). Additionally, Compton-thick sources have very faint emission at 0.1-10 keV (restframe) due to absorbed emission by the circumnuclear clouds of gas and dust. Therefore, QSO2s typically have similar relations between X-ray, optical, and IR as an extremely obscured AGN \citep{2007A&A...467...73T}. 

These sources also showcase a variety of unique features, with frequent tidal features from mergers or interactions \cite[e.g][]{10.1111/j.1365-2966.2012.20652.x,2012MNRAS.426..276B,2023MNRAS.522.1736P}, intense star-formation activity \citep[e.g.][]{2007AAS...210.0215L,2008AJ....136.1607Z,2009ApJ...706..508H}, ionised gas outflows \citep[][]{2010MNRAS.408L...1H,2016MNRAS.460..130V}, and polarised emission with an obscured non-thermal continuum \citep[e.g.][]{2001A&A...366....7V,2005AJ....129.1212Z}. Thus, multi-wavelength observations remain crucial for the validation of the nature of this type of source \citep[e.g. ][]{2002ApJ...568...71S,2003A&A...406..555D,2004AJ....128.1002Z,2016ApJ...831..145Y,2022ApJS..259...18M}

In the `redshift desert' (i.e. $1 \leq z \leq 2$), most of the prominent emission lines in QSO2 sources (e.g. CIV$\lambda1549$; [OIII]$\lambda\lambda4959,5008$) fall either outside the wavelength range of the SDSS optical spectra or in particularly noisy wavelength ranges \citep{2003AJ....126.2125Z,2004AJ....128.1002Z,2013MNRAS.435.3306A,2014AAS...22311504R}. This intermediate redshift regime is still only scarcely explored. For example, in ultraluminous infrared galaxies, QSO2s (ULIRG, L$_{\rm{IR}} \geq 10^{12} L_{\odot}$) have been detected at z\textless 1 and z $\geq 2$ \citep[e.g.][]{2004AJ....128.1002Z,2008AJ....136.2373R,2014A&A...565A..19R}. Studying QSO2s in this redshift range will provide new insights into the properties of AGNs and their host galaxies near the peak of star-formation rate in the universe. The potential correlation between star-formation rate, merger history, and QSO2 sources makes this type of study essential in order to complete our view of the AGN-galaxy co-evolution  \citep{1998AJ....115.2285M,2008ApJ...676...33D,2011ApJ...732....9G,2012MNRAS.426..276B,2015MNRAS.454.4452H, 2017A&ARv..25....2P,2020A&A...634A.116V,2021A&A...650A..84V}.

Following the constant technological innovation in astronomical surveys, the demand to deal with data in a fast and effective manner is not only relevant but also becoming mandatory. Traditionally, classification and physical property estimation has been done using colour-colour methods \citep[e.g. ][]{1956BOTT....2n...8H,2004ApJ...617..746D,2002AJ....123.2945R,2008AJ....136..954G,stern,2004ApJS..154..166L} and spectral fitting codes \citep[template or model-based, e.g.][]{2000A&A...363..476B,2006A&A...457..841I,2018A&A...618C...3G}. While spectral fitting codes have been shown to be extremely useful and to provide solid results, they become difficult to use with massive data sets due to time and computational constraints. In the last decade, the use of machine learning \citep{5392560} in astronomy has increased rapidly, allowing new and more complex models to be developed and helping astronomers retrieve physical meaning from observational data \citep[][]{2019arXiv190407248B}. Machine learning has proved to be particularly efficient at dealing with multidimensional data sets and often gives similar or better results compared to traditional methods \citep[e.g.][]{1993PASP..105.1354O,2004PASP..116..345C,2021A&A...645A..87B, 2023A&A...671A..99E}. The flexibility of this method allows for a wide range of applications, including object classification using photometry data \citep[e.g.][]{2014MNRAS.437..968C,2020A&A...633A.154L,2020A&A...639A..84C,2024A&A...685A.107M,2024AJ....167..169H,2024MNRAS.528.4852P,2024MNRAS.527.4677Z}; morphological classification with images \citep[e.g.][]{2015ApJS..221....8H,2016MNRAS.459..720H,2018MNRAS.476.3661D,2021MNRAS.501.4579B,2022MNRAS.513.1581W,2023MNRAS.519.6149B}; estimation of the redshift of galaxies with tabular and imaging data \citep[e.g.][]{2020A&A...644A..31E,2021MNRAS.507.5034R,2022MNRAS.514....1C,2022A&A...662A..36L,2022MNRAS.512.1696H,2022A&A...666A..87C,2023A&A...671A.153C}; and physical property estimation \citep[e.g.][]{2019A&A...622A.137B, 2021MNRAS.502.2770M,2021ApJ...908...47S,2023MNRAS.520.3529E,2023MNRAS.520..305H}.

Few-shot learning \citep{li2006one, 6630734, wang2020generalizing} is a machine learning method that allows for the construction of models using relatively small data sets. It takes advantage of the size of the data set to reduce computation time and better recognise rare examples or classes. This is achieved by learning the distribution of rare labelled features, contrary to the common case where the main statistical distribution of the features is obtained from a large sample of data. Although this method was first applied to computer vision tasks \citep[e.g.][]{tian2020differentiable,2020arXiv200405805Q,10239698,2023SPIE12718E..0IJ}, it can also be applied to tabular data \citep[e.g.][]{chen2019few,9569130,hegselmann2023tabllm,nam2023stunt,lee2024tableye}. Domain knowledge \citep[e.g.][]{10.1007/978-3-540-87481-2_23,https://doi.org/10.1111/exsy.12019,2022NatSR..12.1040D} plays an important role to be able to use this technique, especially in astronomy. Prior knowledge, from previous observations, and the physical correlations in astronomical data allows the machine learning model to learn in an efficient manner \citep[e.g.][]{2023NatSR..13.1427G}. By combining both techniques, the identification of rare astronomical sources can potentially be enhanced.

Transfer learning in machine learning involves using models pre-trained with specific data sets, either from simulations or real data, to make predictions in new data sets or in different domains \citep[e.g.][]{2022MNRAS.511.1808S,2022MNRAS.509.5657A,2022AAS...24024127K,2022ApJ...929..132V}. Instead of training and optimising a completely new model, transfer learning allows learned patterns and mappings from the original training episode to be re-used for similar tasks. This approach is particularly useful when labelled data for the target task is scarce or expensive to obtain (e.g. spectroscopic data). Therefore, transfer learning enables the application of machine learning to new domains or tasks with smaller data sets, which is particularly relevant in astronomy \citep[e.g.][]{2018JPhCS1085e2014V,2018arXiv181211806K,2019MNRAS.488.3358T}.

Traditional single-colour selection methods of obscured AGNs are often unreliable \citep{2018ARA&A..56..625H}, prompting the exploration of machine learning techniques for more robust selections. Motivated by the difficulty to identify QSO2 sources at the `redshift desert', we have developed a few-shot machine-learning methodology that correlates optical-infrared photometry by linking faint optical emissions and dust features, which are characteristic of obscured AGNs \citep[e.g.][]{2015A&A...574L...9P,2015ApJ...804...27A,2018MNRAS.474.1955H,2023ApJ...950L...5Y,2023MNRAS.522..350I}. Our pipeline is design to handle smaller training data sets ($\sim 200$ sources) and to be applied at the `redshift desert' via transfer learning. To test our pipeline, we performed a blind test, using semi-supervised learning to identify QSO2 candidates. The methodology outlined here offers a photometric pre-selection of QSO2 candidates for multi-wavelength follow-up.

This paper is organised as follows. We briefly describe the photometric data used to distinguish QSO2s from galaxies and the pre-processing steps necessary for the machine learning pipeline in Sect.\ref{section:cp2_data}. The \texttt{AMELIA} pipeline is described in Sect.\ref{section:cp3_meth}, and the statistical metrics to evaluate the pipeline are in Sect.\ref{metrics}. A proof-of-concept task and a semi-supervised approach to identify QSO2s are presented in Sect. \ref{section:Results}. In Sect. \ref{section:Test_recovery}, we test our ability to recover QSO2 candidates through spectroscopic analysis, X-ray and radio cross-matching, and comparison with AGN photometric criteria in the literature. Then, in Sect.\ref{section:pp_estimation}, we derive the physical properties of the QSO2 candidate sample using photoionisation models and spectral energy distribution analysis. Finally, in Sect. \ref{section:context_qso2sample}, we discuss the nature of our sample and how it can be placed into the AGN-galaxy co-evolution scheme. Last, we summarise our results in Sect.\ref{ch6_conc}. Throughout this paper, we adopt a flat-universe cosmology with $H_0$ = 70 km s$^{-1}$ Mpc$^{-1}$, $\Omega_M = 0.3$ and $\Omega_{\Lambda} = 0.7$ \citep{2020A&A...641A...6P}.


\section{Sample and data}
\label{section:cp2_data}

\subsection{Photometric data}
\label{section:phot_data}

\subsubsection{Quasar sample}
\label{qso_data}
Our methodology requires a sample from which the defining characteristics of QSO2 may be learnt. 
For this purpose, we use the 144 sources identified and described by \cite{2013MNRAS.435.3306A} as `class A' QSO2 candidates, which the authors selected from BOSS in SDSS DR9 \citep{2012ApJS..203...21A}. \cite{2013MNRAS.435.3306A} used a selection approach that combined the properties of the emission line and continuum properties: They required a detection of Ly$\alpha \lambda1216$ and CIV $\lambda1550$ at 5$\sigma$ significance or higher, and an emission line full-width half-maximum (FWHM) $\leq$ 2000 km s$^{-1}$. The final sources have narrow emission lines, no associated absorption and weak continuum. The authors discarded narrow-line Seyfert 1 galaxies and broad-absorption-line quasars, since they are Type 1 objects. The selected QSO2 candidates have a redshift between 2 and 4, and a median FWHM of the CIV $\lambda1550$ emission line equal to 1260 km s$^{-1}$. (For more details, see \cite{2013MNRAS.435.3306A}.)

\subsubsection{Galaxy sample}
\label{galaxy_data}
In addition, we select a sample of sources previously classified as galaxies in the SDSS DR16 catalogue \citep{SDSS16} with spectroscopic redshifts, to be used as a `test set' for models.
Our selection criteria are designed to yield galaxies with SDSS magnitudes in a similar range to those of the QSO2 sample above, but without any specific constraints on their colours or morphology. Specifically, our magnitude selection criteria are: (i) $ 19 \leq u \leq 26$; (ii) $ 19 \leq g \leq 24$; (iii) $ 19 \leq r \leq 24$; (iv) $ 19 \leq i \leq 24$; (v) $ 19 \leq z \leq 25$ (AB magnitudes). An SQL query on CasJobs\footnote{\url{https://skyserver.sdss.org/CasJobs/}} was performed using these criteria. We opted to use the aperture-matched modelMag magnitudes, since they are expected to be more reliable than cModelMag, for example, due to the higher signal-to-noise of the resulting colours \citep{2002AJ....123..485S}.  Since we are interested in the `redshift desert', we restrict the galaxy redshift to z \textgreater 1 and z \textless 2 with z=2 being the maximum value for sources classified as a galaxy in SDSS.

When searching for QSO2 in UV-optical photometric surveys, the primary challenge lies in distinguishing them from non-AGN galaxies, given the optical depth of dust and the faint emission diluted by star formation \citep[e.g.][]{1994ApJ...429..582C,2003ApJ...598.1017D}. The uncertainty in the photometric classification boundary, caused by dust and gas obscuration, introduces the possibility of "hidden" QSO2 sources in archival data, which can be untangled by adding infrared information \citep{1988BAAS...20.1067K}. Therefore, we include Wide-field Infrared Survey Explorer \citep[WISE;][]{2010AJ....140.1868W} $W1$, $W2$, $W3$, and $W4$ (3.4, 4.6, 12, and 22 $\mu m$) photometry. These data were obtained by including a cross-match with the AllWISE catalogue in our SQL query. The final galaxy sample is composed of $21,943$ sources.

\subsection{Spectroscopic data}
\label{section:spec_SDSS_data}

We also make use of SDSS BOSS spectra when required for the validation of our machine learning results. We used the SDSS Science Archive Server\footnote{\url{https://dr16.sdss.org/optical/}} to retrieve FITS spectra. The \texttt{zWARNING} flag is used as a quality filter: Only sources with \texttt{zWARNING}$=0$, no problems detected, or $4$, \texttt{MANY$\_$OUTLIERS}, are considered. 

In the `redshift desert', the [OII]$\lambda\lambda3727,3730$ emission is used to determine the redshift of the source. This makes the redshift determinations more prone to errors. Therefore, we will consider the spectroscopic redshift provided by SDSS with caution (more details are provided in Sect.\ref{spec_analysis_QSO2_cand}). It is also important to mention that the spectroscopic data available is close to the limit that can be achieved with a 2.5m telescope and faint sources (r magnitude $\sim 21$). Faint emission lines becomes more difficult to detect as they may be diluted in the noisy continuum.


\section{The \texttt{AMELIA} pipeline}
\label{section:cp3_meth}
\texttt{AMELIA} \footnote{\url{https://github.com/pedro-acunha/AMELIA}} is an interactive, modular, and adaptive machine learning pipeline capable of performing classification tasks using tabular data from astronomy surveys. It takes advantage of six different classifications algorithms and an ensemble method, generalised stacking \citep{WOLPERT1992241}, to combine the outputs using soft-vote and meta-learner correction \citep{10.1093mnrasstw1454,2022A&A...666A..87C, 2023A&A...671A..99E}. The following subsections describe the technical details of \texttt{AMELIA}. In Sect. \ref{section:feat_eng}, we describe data pre-processing tasks required to extract the most out of the available data. Then, in Sect. \ref{section:ML_description} the machine learning methodology is described. In particular, in Sect. \ref{section:amelia_overview} a brief overview of the inner works of our methodology is described.

\subsection{Feature engineering}
\label{section:feat_eng}

\subsubsection{Feature creation}

As the input features for all the models, we used both magnitudes and colours. Due to the diversity of machine learning models, we perform feature scaling \citep{Wan_2019} using classes from \texttt{Scikit-Learn} \citep{scikit-learn}. For all algorithms, the \texttt{RobustScaler} was applied, which ensures that potential outliers in the data are kept. All unique permutations of broadband colours were calculated from the following features: modelMag (i.e. $u-g$, ..., $i-z$); WISE (i.e. $W1-W2$, ..., $W3-W4$); and optical-infrared colours (i.e. $u-W1$,..., $z-W4$).  When we perform binary classification tasks, a binary (0/1) target feature is created, where 0 indicates a normal galaxy and 1 indicates a QSO2.

\subsubsection{Imputation}
In astronomy, missing values are commonly due to instrumental limitations (i.e. non-detections) and do not necessarily represent an incomplete data set, as they may carry a relevant physical meaning. For instance, some galaxies have sudden spectral breaks at specific wavelengths that can manifest themselves as "dropouts", revealing both the source's nature and the redshift. Nonetheless, missing values can cause problems on how different algorithms interplay and digest data. A common solution is imputation, where missing values are replaced with an alternative value with statistical or physical meaning.

The effects of imputation have also been extensively studied in \citet{2023A&A...671A..99E}. The authors explored the impact that different imputation techniques, such as the mean, median, minimum and a so-called "magic value" (e.g. -99.9), have in decision-tree and distance based, deep learning and meta-learner ensemble models. The chosen imputation techniques impact the model's performance and should therefore be selected according to the scientific problem.

For the QSO2 sources from \cite{2013MNRAS.435.3306A}, missing values are present in the WISE photometry for 74 sources, which represents $\sim 51\%$ of the sample (identified using a $2\arcsec$ cone search). To impute the missing values, we used the \texttt{KNNImputer}, from \texttt{Scikit-Learn, version 1.2.1}, with the following configuration: \texttt{missing$\_$values}=0; \texttt{n$\_$neigbors}=6; and \texttt{weights}= \texttt{distance}. As in the k-nearest neighbours algorithm, this method predicts the missing value by computing the mean value based on the distance between points in the training set and the number of defined neighbours.

\subsection{Machine learning}
\label{section:ML_description}
After the feature engineering process, described in Sect.\ref{section:feat_eng}, the data are divided into training and test set, depending on the final objective (for further details, see Sect.\ref{section:Results}). Every time a data set is split, the procedure is performed randomly, never considering the balance between the properties of the features. This allows experiments in the stacking process to be applied and compared with the each base learner. However, this variability can easily be changed and different training-test splits tested.
\begin{figure*}
    \centering
    \includegraphics[width=\linewidth]{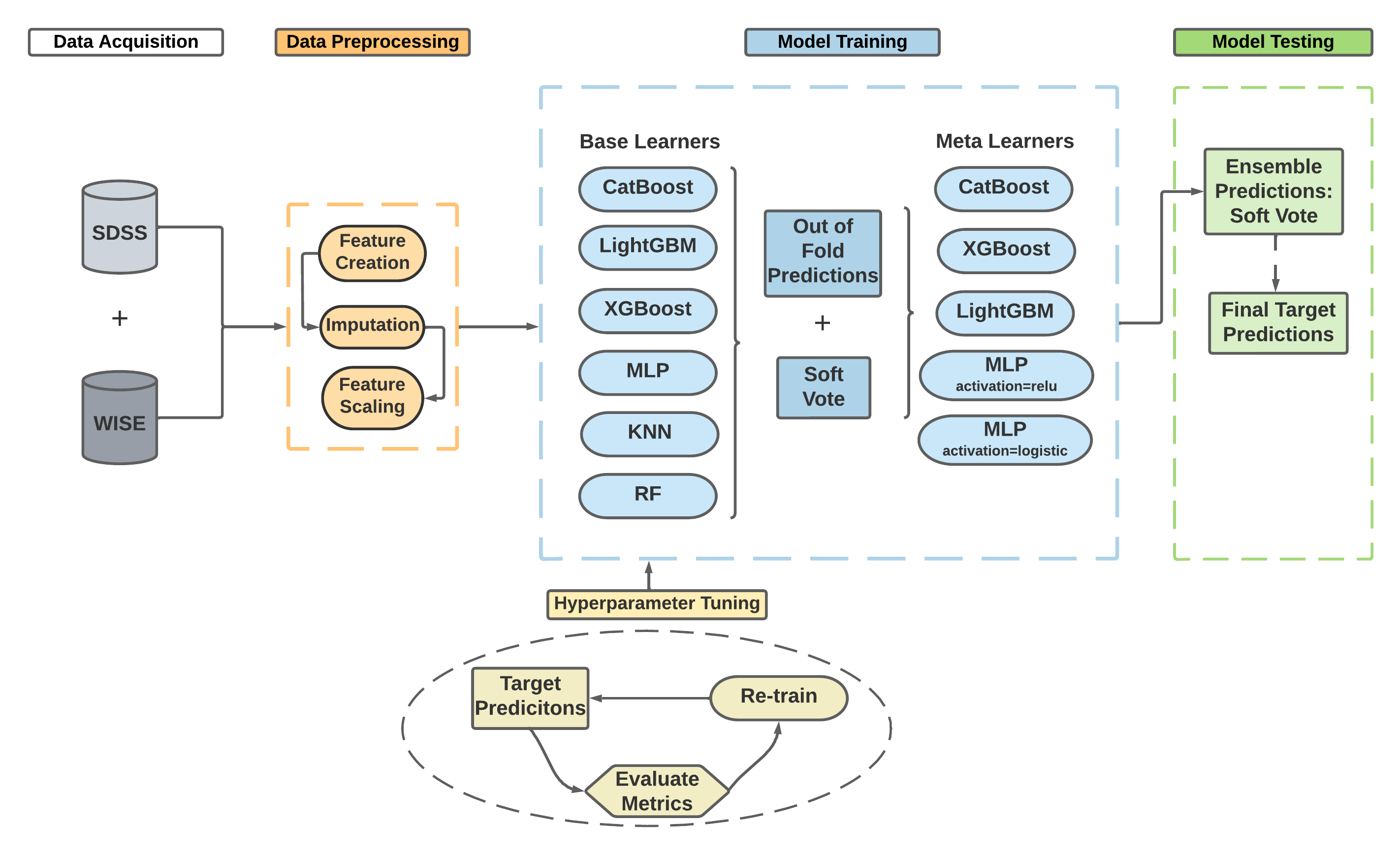}
    \caption{Flow diagram for the \texttt{AMELIA} pipeline. The process begins with the assembly of SDSS and WISE data. As seen in the orange box, feature creation is used to generate optical, optical-IR, and IR colours. The imputation method is applied to the training set and followed by feature scaling due to the use of different algorithms. The blue box shows a description of the model training procedures, but we recount them here as well: Each base learner is first loosely optimised for generalisation. Subsequently, out-of-fold predictions are combined with the average predictions from the base learners, utilising a soft-vote mechanism for the training of the meta-learners. The predictions from the meta-learners are then combined through ensembling, using a soft-vote fusion that results in the final predictions for the test set.}
    \label{fig:fc_amelia}
\end{figure*}

In Fig.\ref{fig:fc_amelia}, the flow diagram of the \texttt{AMELIA} pipeline is shown. Six different algorithms were used to perform a binary classification, which we call base learners. We combine one distance-based algorithm, four decision tree-based algorithms, and a deep learning-based model (see more details in Sect.\ref{section:ML_algorithms}).

\subsubsection{\texttt{AMELIA} overview}
\label{section:amelia_overview}

In our pipeline, we explored two distinct approaches: supervised and semi-supervised learning. In the supervised learning approach, we adhered to the conventional procedure, where the data set is divided into training and testing sets, utilising k-fold cross-validation, with the class labels of all sources known.

In contrast, with the semi-supervised learning approach, we adopted a unique strategy where the test set consists exclusively of a single class, that is, galaxy class, \citep[see][ for an application to galaxy morphology classification]{2023RAA....23k5019J}. Our approach is based on the foundational one-class classification problem \citep{MOYA1996463,NIPS1999_8725fb77}, taking advantage of the capabilities of one-class classifiers to discern distinct patterns within a multidimensional space. These classifiers are trained using a set of positive examples \citep[e.g.][]{tax2001uniform,yu2002pebl}. To augment the generalisation capability of our models, we embraced a semi-supervised one-class approach \citep[e.g][]{2017arXiv170500797B,akcay2019ganomaly,CAI2023102794}, introducing a limited number of contrasting examples during training and applying to a single class test set. Therefore, while the model is trained with two distinct classes, when the model is applied to the one-class test set, the main target is to identify the galaxy class. Due to the contrasting examples, we then consider sources that are not classified as galaxy to be QSO2 candidates.

In our approach, the model is trained with both QSO2 sources and a subset of SDSS galaxies, ensuring a balanced training set, that is, same number of examples for both classes. Subsequently, the model is applied to the one-class test set (i.e. galaxies) to compute the probabilities for the predictive class (i.e. QSO2). Then, a classification threshold is set to refine the predictions. Specifically, this semi-supervised one-class approach identifies sources labelled as galaxies in the SDSS spectroscopic data but exhibiting photometric characteristics consistent with a QSO2. Our proof of concept uses data sets with diverse redshift ranges, detailed in Sect.\ref{section:cp2_data}, resulting from specific selection criteria in the chosen data set. Given the absence of a catalogue of spectroscopically confirmed QSO2 in the `redshift desert', a redshift transfer-learning method is employed by leveraging prior domain knowledge concerning the correlation between optical and infrared photometric properties from QSO2 at $z \geq 2$.

\subsubsection{Algorithms}
\label{section:ML_algorithms}
AMELIA takes advantage of the predictive capabilities of three decision tree-based algorithms, all falling under the category of gradient-boosting decision tree algorithms. This choice is intentional, as these algorithms demonstrate superior performance and exhibit inherent distinctions in their underlying mechanisms\footnote{A benchmark study between the three algorithms was performed by \citet{2018arXiv180904559A}. Additionally, an interesting analysis of why tree ensemble models can outperform deep learning models with tabular data can be seen in \citet{shwartz-ziv2021tabular}.}. The nuances between the three gradient-boosting decision tree algorithms, namely \texttt{XGBoost (version 1.7.3)}\footnote{\url{https://XGBoost.ai/}} \citep{chen2016xgboost}, \texttt{CatBoost (version 1.1)}\footnote{\url{https://catboost.ai//}} \citep{prokhorenkova2018catboost}, and \texttt{LightGBM (version 3.3.3)}\footnote{\url{https://lightgbm.readthedocs.io/}} \citep{Ke2017LightGBMAH} are described in \citet{2022A&A...666A..87C} and \citet{2023A&A...671A..99E}.

For completeness and diversity, we incorporate a k-nearest neighbours algorithm, a randomised decision trees algorithm, and a multi-layer Perceptron classifier, \texttt{MLPClassifier}, from \texttt{Scikit-Learn (version 1.1.3)}\footnote{\url{https://scikit-learn.org/}}. The inclusion of a diverse set of algorithms is crucial as it enables a wider range of predictions. This diversity will become particularly pertinent later within the context of the generalised stacking approach with meta-learning correction (see Sect.\ref{section:generalised_stack}).

In Table \ref{tab:base_meta_alg}, the hyper-parameters of the algorithms used as base learners and meta-learners are described. The number of estimators were defined as being the minimum number to obtain reliable metrics, as increasing the estimators can lead to overfitting. This was done using a randomised grid search for \texttt{XGBoostClassifier} and replicated for the remaining algorithms. For the \texttt{MLPClassifier}, we set the initial hidden layer with the same number of layers as the \texttt{n\_estimators}. Additional layers were added and a manual optimisation was performed. For the \texttt{KNeighborsClassifier}, we set a range of \texttt{n\_neighbours} from 0 to 20, and selected the one that had the best metrics. 
\begin{table}
\begin{center}
\caption{Description of the algorithms and hyper-parameters used as base learners and meta-learners.}

\begin{tabular}{c c}
\hline
Algorithms    & Hyper-parameters                                                                \\ \hline
\texttt{KNN}           & n\_neighbours = 12                                                              \\ \hline
\texttt{Random Forest}  & \begin{tabular}[c]{@{}c@{}}\texttt{n\_estimators} = 50\\ \texttt{max\_depth} = None\end{tabular} \\ \hline
\texttt{XGBoost}        & \begin{tabular}[c]{@{}c@{}}\texttt{n\_estimators} = 50\\ \texttt{max\_depth} = 6\end{tabular}    \\ \hline
\texttt{CatBoost}       & \begin{tabular}[c]{@{}c@{}}\texttt{n\_estimators} = 50\\ \texttt{max\_depth} = 6\end{tabular}    \\ \hline
\texttt{LightGBM}       & \begin{tabular}[c]{@{}c@{}}\texttt{n\_estimators} = 50\\ \texttt{max\_depth} = -1\end{tabular}   \\ \hline
\texttt{MLP}            & \begin{tabular}[c]{@{}c@{}}\texttt{Hidden layer sizes} = (50,100,50) \\ \texttt{activation} = relu\end{tabular}\\ \hline
\end{tabular}
\label{tab:base_meta_alg}
\end{center}
\end{table}

\subsubsection{K-fold cross-validation}

To assess the performance of each base learner, we employed a data partitioning strategy known as k-fold cross-validation \citep[refer to, for instance, ][]{Hastie2009}. In this methodology, the data set is randomly divided into five folds, ensuring the preservation of both the number of sources and the target class distribution. Following this division, the models are trained on the partitioned data set, utilising k-1 folds for training, while the remaining fold is designated for validating the predictions, known as out-of-fold (OOF) predictions. This iterative process repeats through the defined k iterations and an average is calculated on the predictions for the final metric analysis.

K-fold cross-validation proves to be an effective methodology to prevent overfitting, especially in smaller data sets where the conventional training-test-validation split may not be feasible \citep{b0094eaa-462f-39e3-8d02-e2fda45ff061,kfold,luntz_kfold,wang_kfold}. An additional advantage is to obtain OOF predictions, even for the training set. In this study, the k-fold cross-validation method plays a crucial role in validating the performance of each individual learner before implementing the stacking approach.

\subsubsection{Approach to generalised stacking}
\label{section:generalised_stack}
Our approach to handling the output predictions of the base learners closely follows the
generalised stacking methodology presented by \citet{2023A&A...671A..99E} and \citet{2022A&A...666A..87C}. The primary difference is how the final combination of the predictions is performed. In this study, we opt for a soft vote, that is, average vote, coupled with a meta-learner approach to combine the probabilistic predictions. 

Upon receiving predictions from the six base learners for the test set, a new data set is constructed, comprising individual predictions from each base learner. Additionally, a new feature is created by averaging these predictions. To integrate the predictions of the individual learners and the average feature, we employ a combination of five meta-learners: \texttt{XGBoostClassifier}; \texttt{CatBoostClassifier}; \texttt{LGBMClassifier}; \texttt{MLPClassifier} (as described in Table \ref{tab:base_meta_alg}); together with an additional \texttt{MLPClassifier} with \texttt{activation=logistic}. The inclusion of the \texttt{MLPClassifier} model with a distinct activation model aims to increase the diversity of predictions by adding an additional deep learning based model.

The final predictions result from an ensemble of the individual outputs of the meta-learners, utilising a soft vote to generate the final classifications. This methodology explores how various algorithms analyse the distribution of predictions and combine them into a robust classifier. Furthermore, the exclusion of the Random Forest and KNN algorithms is attributed to the influence of outliers on the number of outputs, which can impact the soft vote combination. This decision was validated through a supervised approach, as detailed in Sect.\ref{section:sc_high_z}.

In summary, one can interpret our approach as an analogous of the ensemble random forest, where instead of randomly decision-trees, we have individual base models. Then, instead of combining the predictions for each model, we use a meta-learner to analyse the variance between the base learners' predictions, adding the average predictions as a feature. The main advantage of this method is that the output model will be robust to outliers by studying the variance between predictions and preventing systemic biases underlying particular models. Finally, to avoid selection bias from a single meta-learner, we average the predictions from four meta-learners. With this approach, one can create more robust predictions by leveraging the statistical validation provided by the ensemble models.

\section{Statistical metrics}
\label{metrics}

Statistical metrics quantify the ability of a particular model to retrieve the desired information \citep[e.g.][]{ml_metrics}. Herein, we describe the statistical metrics used to validate the perform of our machine learning methodology.

Accuracy is the fraction of correct predictions for all classes compared to the to the total number of predictions:

\begin{equation}
    \rm \textit{A} = \frac{TP +TN}{TP + TN + FP + FN}\,.
\end{equation}
\noindent Here, TP is the number of true positives, TN is the number of true negatives, FP is the number of false positives, and FN is the number of false negatives for a given class.

 Precision, or purity, is the fraction of correct predictions for a given class compared to the overall number of predictions for that class:

\begin{equation}
\rm P = \frac{\rm TP}{\rm TP + \rm FP}\,.
\end{equation}
\noindent Unlike the accuracy metric, precision focus on the positive predictions to asses their quality.

Recall, or completeness, is the fraction of correct predictions for a given class compared to the overall number of positive cases for that class:

\begin{equation}
\rm R = \frac{\rm TP}{\rm TP + \rm FN}\,,
\end{equation}
\noindent where FN is the number of false negatives in each class. The symbiotic relation between precision and recall makes their use valuable to asses the quality of the predictions.

By combining the precision and recall metrics, one can create informative metrics such as the F1-score \citep[e.g.][]{F1score}. The F1-score is the harmonic mean of the precision and recall,

\begin{equation}
\rm F1 = 2 \frac{P \cdot R}{P+R},
\end{equation}
\noindent where equal weight is given to precision and recall.

Specificity, also called true negative rate, computes the true negative fraction identified by the model:
\begin{equation}
\rm specificity = \frac{\rm TN}{\rm TN + \rm FP}\,.
\end{equation}
High specificity implies that a high fraction of true negatives is recognised. 
\section{Results from the application of \texttt{AMELIA}}
\label{section:Results}

\subsection{Proof of concept for QSO2 selection}
\label{section:sc_high_z}

Before selecting new QSO2 candidates, we first conduct a test to ascertain the performance of our pipeline for selection of QSO2 in a fully labelled data set. By employing a supervised classification on SDSS-WISE photometric data, we assess the reliability of the binary classification. Our primary goal is to establish a control test to explore correlations in optical-infrared photometric data.

\begin{table*}
\begin{center}
\caption{Classification evaluation metrics for \texttt{KNNeigbors}, \texttt{RandomForest}, \texttt{LightGBM}, \texttt{CatBoost}, \texttt{XGBoost}, \texttt{MLP}, and generalised stacking with a threshold $\geq 0.5$.}
\begin{tabular}{@{}*{6}{c}@{}}
\hline
Algorithm & Accuracy  &  Precision & Recall & F1-Score & F1-Score, cv=10\\
\hline
\texttt{KNNeigbors} & 0.879  & 0.897 & 0.887 & 0.879 & 0.876 \textpm 0.063\\
\texttt{RandomForest} & 0.879 & 0.881 & 0.882 & 0.879 & 0.867 \textpm 0.080\\
\texttt{XGBoost} & 0.966 & 0.965 & 0.965 & 0.965 & 0.943 \textpm 0.041\\
\texttt{CatBoost} & 0.966  & 0.966 & 0.968 & 0.965 & 0.926 \textpm 0.041\\
\texttt{LightGBM} & 0.948 & 0.949 &  0.949 & 0.948 & 0.921 \textpm 0.046\\
\texttt{MLP} & 0.948 & 0.950 & 0.952 & 0.948 & 0.921 \textpm 0.035\\
Generalised stacking &  0.966 & 0.966 & 0.968 & 0.965 & --\\
\hline
\end{tabular}
\tablefoot{The final column shows the F1-Score metric using k-fold cross-validation, with k=10, that allows the standard deviation for the final classifications.}
\label{table:sdss_clf_eval}
\end{center}
\end{table*}

For this proof of concept, the data set was constructed using $144$ sources randomly selected from our SDSS galaxy sample (see Sect.\ref{galaxy_data}), and all $144$ spectroscopically confirmed sources in our QSO2 sample (see Sect. \ref{qso_data})\footnote{An examination of how the class distribution of the training data impacts the overall performance of the models is presented in Sect.\ref{section:training_dist}.}. After combining the data from the two classes, we performed a train test splif using $70\%$ of the total data for training and $30\%$ of the total data for testing. 

The \texttt{AMELIA} pipeline was able to successful distinguish between galaxies and the QSO2, validating the application of machine learning for this task. Detailed in Table \ref{table:sdss_clf_eval} are the statistical metrics for each individual model, together with the metrics obtained after generalised stacking with a threshold $\geq 0.5$. Individual models exhibited satisfactory performance, with a slight advantage noted for the decision-tree-based models, \texttt{XGBoost} and \texttt{CatBoost}, compared to the rest of the models. 

The metrics for the generalised stacking, also seen in Table \ref{table:sdss_clf_eval}, may not be sufficiently improved to validate the advantage of generalised stacking in this context. Methodological intricacies arise when models yield similar results, causing meta-learners to mimic clustered values for the final predictions, converging around the optimal value. However, this does not diminish the potential of the methodology to identify QSO2 sources, where individual learners achieve an F1-score $\geq 0.88$. As the size of the data set increases, which would be the case when our methodology is applied to future survey data, we anticipate an improvement in performance \citep[as seen in ][]{2023A&A...671A..99E,2022A&A...666A..87C}.

\subsection{Selection of new QSO2 candidates}
\label{section:semi_supervised}

Having shown in Sect. \ref{section:sc_high_z} that \texttt{AMELIA} is capable of successfully identifying QSO2, we now proceed to apply the pipeline to the task of identifying new QSO2 candidates. In particular, the objective is to select QSO2 that were previously misclassified in SDSS as galaxies, after template fitting. Thus, we now use the \texttt{AMELIA} pipeline in semi-supervised mode (see Sect. \ref{section:amelia_overview}), utilising a balanced training set (same number of sources per class), and a classification probability threshold $\geq 0.8$. By using this probability threshold instead of the often used `default' value 0.5, we aim to select sources with a higher confidence of being a QSO2 \citep[see][ for different study cases with classification threshold tuning and Appendix \ref{section:training_dist} for a more detailed analysis]{2021A&A...645A..87B,2022MNRAS.516.4716A,2023RAA....23k5019J}.

For this task, the training set contains the $144$ spectroscopically confirmed QSO2, and a further $144$ sources selected randomly from our SDSS galaxy sample, so make up the balanced training set. The testing set was created using the remaining $\ 21\,799$ sources from the galaxy sample; no spectroscopically confirmed QSO2 were included in this set. 

When applying the \texttt{AMELIA} pipeline on the test set, 366 sources were 
classified as QSO2 with a classification probability of $\geq 0.8$. In the following sections, we will address the reliability of our new QSO2 candidates by analysing photometry and spectra, including multi-wavelength information where available.

\section{Testing the recovery of QSO2}
\label{section:Test_recovery}

In this section, we assess the robustness of our new QSO2 candidate sample by examining available optical photometry and spectra, and X-ray and radio detections. This procedure ensures the reliability of the identified candidates; while not providing a definitive classification of the target sample as QSO2 sources, with multi-wavelength data being essential. Nonetheless, the identification of specific fingerprints indicative of AGN activity significantly enhances our confidence in the efficacy of our selection method.

\subsection{Searching for AGN signatures in the spectrum}
\label{spec_analysis_QSO2_cand}

Of the $366$ candidates, only $138$ had clean spectra in SDSS DR16 (i.e. without any problematic flag, \texttt{zwarning}=0). Thus, our spectroscopic analysis includes only these $138$ candidates.

 After downloading the SDSS spectra, we considered the spectroscopic redshifts (described in Sect. \ref{section:spec_SDSS_data}) to search for AGN associated emission lines. We found that the redshift listed in the SDSS database occasionally did not match what we obtained from the observed emission lines \citep[e.g.][ detected 1900 broad-line quasars in SDSS DR16, around 0.002$\%$ of their sample, had catastrophically wrong redshifts]{2022ApJS..263...42W}. In such cases, we adopted our new value of the redshift for the source. An example is provided in Appendix \ref{appendix:source_wrongz}. When no clear emission lines were detected, we could not validate the provided spectroscopic redshift and therefore assumed the provided redshift value.
 
 Our main objective was to identify [NeV]$\lambda3426$, successfully used as an AGN identification criterion \citep[e.g.][]{1998A&A...329..495S, 2010A&A...519A..92G, 2013A&A...556A..29M, 2018A&A...620A.193V}, and [NeIII]$\lambda3869$ emission lines. [NeV]$\lambda3426$ is a high ionisation line with a relatively high ionisation potential (IP; Ne$^{4+}$ IP=$97.16$ eV) requiring a hard ionising source, such as an accretion disk of an AGN \citep[][]{2001ApJ...549L.151H}. %
 \citep[e.g.][]{1998A&A...329..495S, 2010A&A...519A..92G,2013A&A...556A..29M,2016MNRAS.456.3354F}. Alternative, other more "exotic" ionisation sources might be able to produce [NeV]$\lambda3426$ such as supernovae, Population III stars, Wolf–Rayet stars, stripped stars in binaries, high-mass X-ray binaries, and hot, low-mass evolved stars \citep[e.g.][]{2011MNRAS.415.2182F,2017PASA...34...58E,2019A&A...622L..10S, 2023MNRAS.525.2916M}.

In our spectroscopic analysis, we fitted each emission line and identified $30$ [NeV]$\lambda3426$ and $52$ [NeIII]$\lambda3869$ emitters with S/N $\geq 3\sigma$. In total, we found $65$ objects with [NeV]$\lambda3426$ or [NeIII]$\lambda3869$ emission lines, around $47\%$ of the total number of candidates.

\begin{figure*}
\centering
\includegraphics[width=0.8\linewidth]{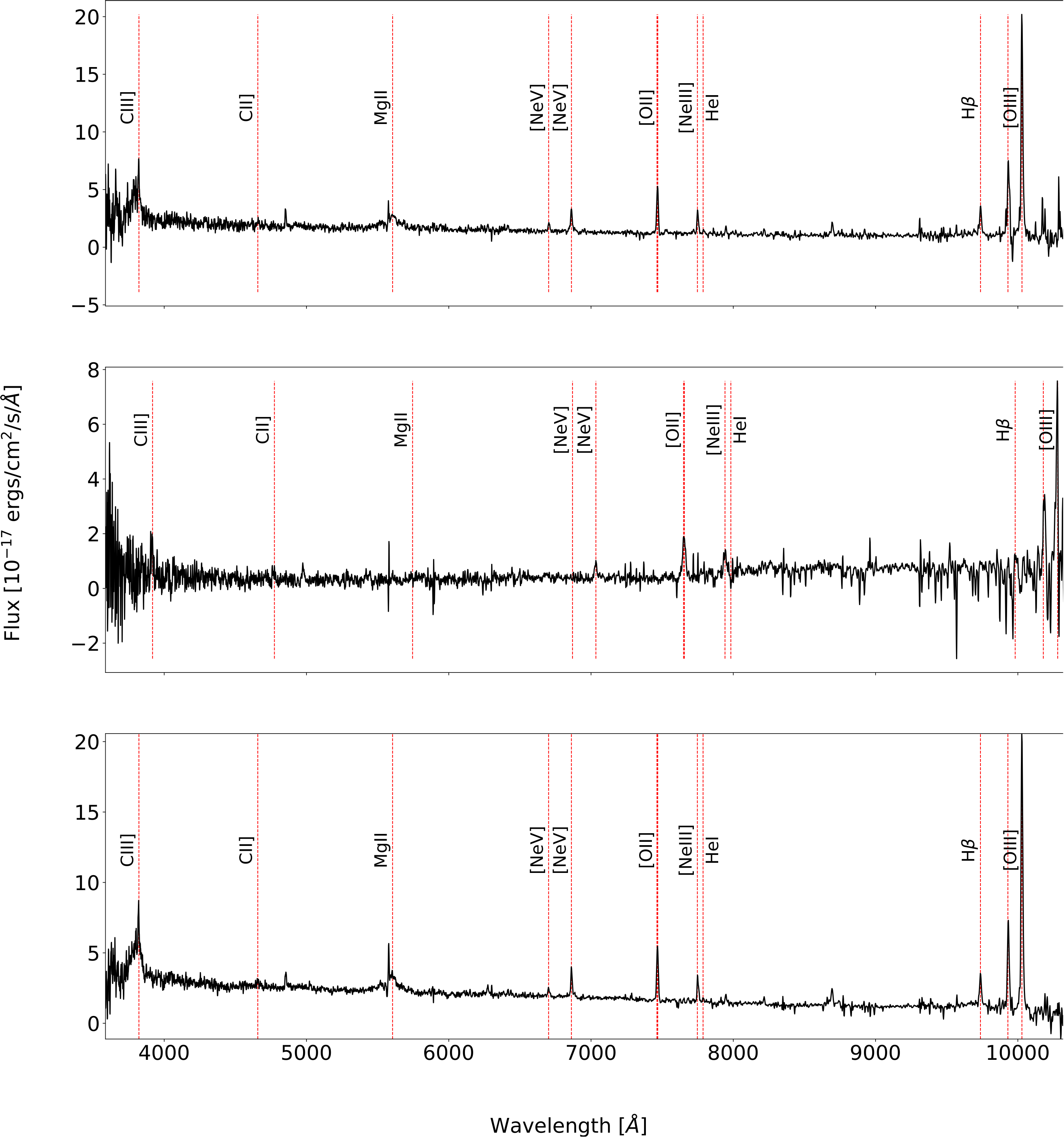}
\caption{SDSS spectra from BOSS spectrograph for a sample of three QSO2 candidates with a [NeV]$\lambda3426$ detection. From top to bottom: SDSS J141542.16+522206.9, SDSS J013056.89-022638.1, and SDSS J141542.16+522206.9. The red vertical lines indicate where the different emission lines should be with respect to the SDSS spectroscopic redshift. }
\label{fig:spectra_candidates}
\end{figure*}

Figure \ref{fig:spectra_candidates} shows the spectrum of three [NeV] emitters from our sample of QSO2 candidates. The final candidates were divided into three groups based on the detected emission lines \footnote{Throughout this paper, the sources with the emission lines will be name as follows: [NeV] emitters for [NeV]$\lambda3426$ emitters; and [NeIII] emitters for [NeIII]$\lambda3869$ emitters.}: Clear AGN: [NeV]$\lambda3426$ detected with S/N $\geq 3\sigma$; Ambiguous AGN: [NeIII]$\lambda3869$ detected with S/N $\geq 3\sigma$; Unclear: No [NeV]$\lambda3426$ or [NeIII]$\lambda3869$ detected.

\begin{figure}
    \centering
    \includegraphics[width=1\linewidth]{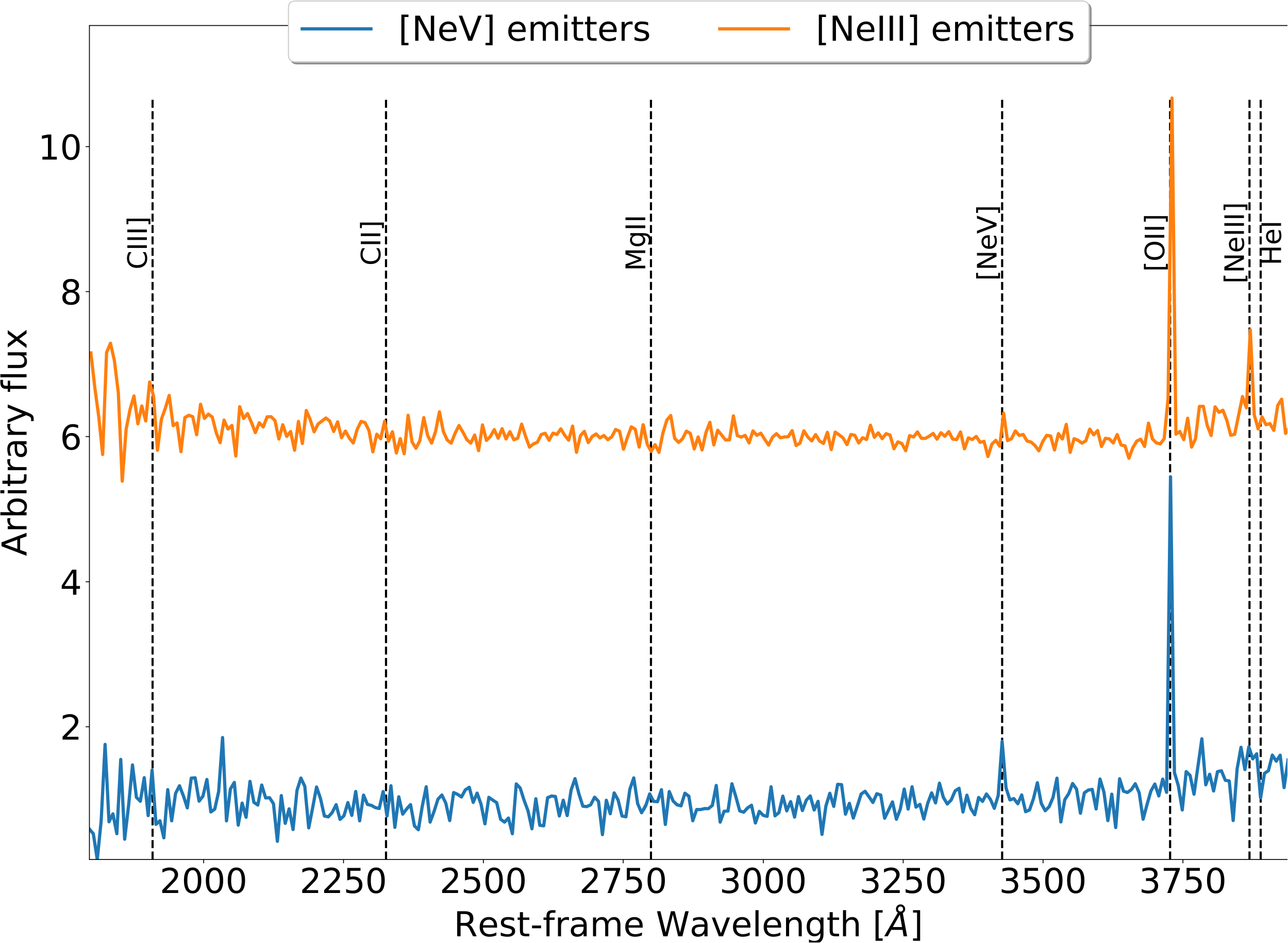}
    \caption{Stacked spectra of the [NeV] and [NeIII] emitters from the QSO2 candidate sample. The fluxes for the [NeIII] emitters is offset for a better visualisation.}
    \label{fig:stacked_spec}
\end{figure}

Fig. \ref{fig:stacked_spec} shows the stacked spectra for the [NeV] and [NeIII] emitters. For stacking the spectra, we used the software \texttt{specstack}\footnote{\url{https://github.com/astrom-tom/specstack}} \citep{2019ascl.soft04018T}, with a limit of $3\sigma$ clipping and rest-frame limit between 1795 and 3838 \textup{~\AA}. Due to the low S/N of the spectra only the most prominent emission lines are detected, in particular [NeV]$\lambda3426$, [OII]$\lambda\lambda3727,3730$, and [NeIII]$\lambda3869$.

Additionally, we compared our methodology with random selection. From the test set, $366$ sources were randomly selected. From the selected sources, we investigated $255$ sources, with clean spectra, and found that $13$ of them had [NeV]$\lambda3426$ lines. This corresponds to 5$\%$ of the total random sample. As expected, our methodology is much more efficient at selecting QSO2 candidates than a random selection. We argue that our classification pipeline may use specific features, that is, combination of magnitudes and colours with specific weights, within the broadband SED that indicate the presence of an obscured AGN, even in the cases where no clear AGN emission lines could be detected in the SDSS spectrum, most probably due to the low S/N or instrumentation limitations \cite[e.g.][ see references herein]{2013A&A...556A..29M,2018A&A...620A.193V}. This is mainly due to the optically faint nature of the sample and designed strategy of the SDSS facility. Follow-up spectroscopic both in optical and infrared regimes will allow us to improve this analysis and distinguish QSO2 from star-forming galaxies.

An unsupervised learning analysis was also performed to study the possible variability within the photometric data between [NeV] and [NeIII] emitters and non-[Ne] emitters. We applied the \texttt{KMeans} algorithm with PCA dimensionality reduction on the input features to identify potential clusters and structures within the data. The number of clusters was set to three, representing the three different emitter labels defined previously. We found no clear separation or intrinsic property to distinguish between the sources, implying no significant variance exists between our sub-classes. This results supports the hypothesis that the unclear sources might have similar features as clear and ambiguous AGNs, requiring high spectroscopic resolution.

\subsection{Cross-matching with radio and X-ray catalogues}

Given that QSO2 might have faint X-ray emission \citep[e.g.][]{2003AJ....126..632B,2006ApJ...637..147P,2007A&A...467...73T}, the detection of X-ray emission from our sources would corroborate our classification. Additionally, radio emission is present in previously identified QSO2 sources at low redshift \citep[e.g.][]{2014MNRAS.440.3202V,2021A&A...650A..84V}, allowing for better characterisation of the sources.

In this section, we examine currently available X-ray and radio data sets to evaluate the detection of our QSO2 candidate sample. Also, we make use of the Millions of Optical-Radio/X-ray Associations (MORX) Catalogue, v2\footnote{\url{https://cdsarc.cds.unistra.fr/viz-bin/cat/V/158}} \citep{2024OJAp....7E...6F} to cross-match our candidates using the VizieR service.

\subsubsection{X-ray cross-identification}

The coverage of our sources by deep X-ray observations from XMM-Newton and Chandra is patchy and does not lend itself to statistical studies. However, we highlight that XMM-Newton \citep{2012ApJ...756...27L} has detected the five QSO2 candidates: SDSS J092106.72+302737.8, SDSS J014104.76-005209.2, SDSS J020711.42+021549.5, SDSS J141542.16+522206.9, and SDSS J011219.62+023732.7. Additionally, SDSS J160852.36+443028.0 has been detected in the XMM-Newton slew survey \citep{2008A&A...480..611S}. Chandra \citep{2016A&A...593A..55V} has detected three QSO2 candidates, two of which also have XMM-Newton detection: SDSS J141542.16+522206.9, SDSS J092106.72+302737.8, and SDSS J014546.90-043318.8. Additionally, SDSS J084126.29+411059.9 is included in the eROSITA Data Release 1 \citep{2024A&A...682A..34M}. 

While these detections confirm the AGN nature for these 10 cases, we can still be missing Compton-thick sources, as previously observed \citep[e.g.][]{2006ApJ...637..147P,2021ApJ...908..185C,2023ApJ...950..127C}.

\subsubsection{Radio cross-identification}
We cross-matched our sample with the LOw-Frequency ARray (LOFAR) Two-metre Sky Survey \citep[LoTSS,][]{2022A&A...659A...1S} Data Release 2, and obtained a total of 20 detected sources from our sample. Since we get a higher number of sources, we have performed a statistical analysis for this sub sample.

We took into account the following equation from \cite{2024A&A...683A..23D} to calculate the total flux density ($S_\nu$):
\begin{equation}
S_{\nu} = \frac{L_{\nu}(1+z)^{1-\alpha}}{4\pi d_L^2},
\end{equation}
\noindent where $L_\nu$ is the source luminosity, $d_L$ is the luminosity distance, and $(1+z)^{1-\alpha}$ considers the $K$-correction with spectral index $\alpha$. We could rearrange this equation to estimate $L_\nu$, with $S_\nu$ as the total flux density estimated in the LoTSS catalogue. We used this estimation to determine if the origin of the radio emission is star-formation driven or AGN driven. Following \cite{2016MNRAS.462.1910H}, we used the following criterion to set a confidence level on the origin of radio emission: log$_{10}$L$_{\rm{150MHz}} \geq 23.5$ \citep[see also][ in particular, Fig.3 and Table 4 therein]{2012MNRAS.421.1569B}. We estimated the following range for the L$_{\rm{150MHz}}$: log$_{10}$L$_{\rm{150MHz}}=24.32 - 27.05$ W Hz$^{-1}$. Thus, the measured L$_{\rm{150MHz}}$ shows a high probability that radio emission has an AGN origin since it is within the interval mainly populated by detected LOFAR AGNs \citep[][see Fig. 17]{2016MNRAS.462.1910H}.

Finally, by cross-matching with Faint Images of the Radio Sky at Twenty-cm \citep[FIRST,][]{1994ASPC...61..165B} catalogue, we identified five QSO2 candidates: SDSS J083317.63+242113.8, SDSS J020711.42+021549.5, SDSS J115621.12+552723.7, SDSS J000844.13-015212.0, and SDSS J160852.36+443028.0 (also detected with LoTSS and XMM-Newton). From the Rapid ASKAP Continuum Survey \citep[RACS,][]{2021PASA...38...58H}, three QSO2 candidates are detected: SDSS J020711.42+021549.5 (also detected with FIRST and XMM-Newton), SDSS J084035.02+225159.2, and SDSS J224340.32+255931.8 (also detected in LoTSS). Future studies in the radio and X-ray wavelength,  will allow us to better characterise the nature of our candidates.

\subsection{Comparison with AGN colour criteria from the literature}

Since only 38$\%$ of our QSO candidates have clean SDSS spectra, it is interesting to apply colour-colour criteria from the literature to further extend our analysis. Our QSO2 candidate sample was compared with the colour criteria for the selection of AGN sources, including: \citet{stern,2012MNRAS.426.3271M,2016MNRAS.462.2631M,2018MNRAS.478.3056B,2023A&A...679A.101C}.

\begin{figure}
    \centering
    \resizebox{\hsize}{!}{\includegraphics{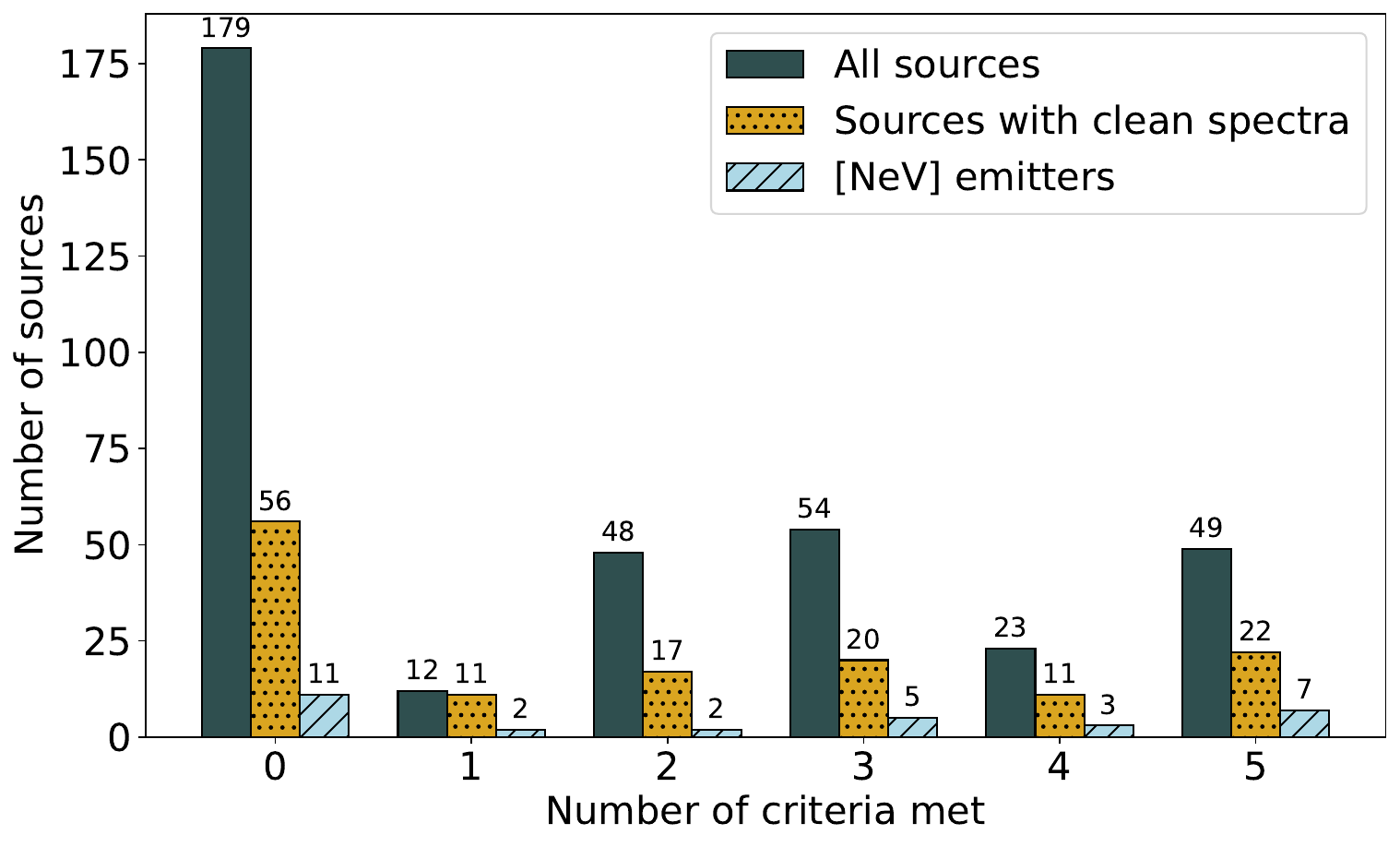}}
    \caption{Bar plot illustrating the performance of the colour criteria in selecting AGN for the QSO2 candidates, as described in the studies by \citet{stern,2012MNRAS.426.3271M,2016MNRAS.462.2631M,2018MNRAS.478.3056B,2023A&A...679A.101C}. Each bar represents a specific subset: (a) [NeV] emitters (30 sources), (b) sources with clean spectra (138 sources), and (c) all candidate sources (365 sources).}
    \label{fig:AGN_colour_barchart}
\end{figure}

Most of the criteria are based on a combination of MIR colours, and the results are summarised in Fig. \ref{fig:AGN_colour_barchart}. Three groups were used in this analysis: (a) [NeV] emitters (or clear AGN; see Sect.\ref{spec_analysis_QSO2_cand}), (b) candidates with clean spectra, and  (c) all QSO2 candidates.

For group (a), the results are as follows: seven sources ($23\%$) met all criteria, 15 sources ($50\%$) met at least three criteria, 11 sources ($\sim 37\%$) did not meet any criteria. On group (b): 22 sources ($\sim 16\%$) met all criteria, 65 sources ($\sim 47\%$) met at least three criteria, and 56 sources ($41\%$) did not met any criteria. Finally, for group (c): 49 sources ($\sim 13\%$) met all criteria, 126 sources ($\sim 35\%$) met at least three criteria, and 179 sources ($\sim 49\%$) did not meet any criteria.

Additionally, we performed the same analysis using the colour selection criteria proposed by \citet{2017ApJ...849...53H}, where a combination of a simple MIR colour ($W1-W2\geq 0.7$; Vega) and an optical-infrared colour [($u$-W3[AB])$\geq$ 1.4(W1-W2[Vega])+3.2], is used to select obscured quasars. For group (a), three candidates ($\sim 10\%$) met this criterion. In group (b), 12 candidates ($\sim 9\%$) satisfied the criterion. Finally, in group (c), 30 candidates ($\sim 8\%$) met the criteria.

Colour criteria serve as an effective but conservative pre-selection tools, which fall short in capturing all relationships across the SED. Relying on two or three highly correlated features only partially taps into the available data potential. For example, the detection percentage of [NeV] emitters surpasses the number obtain using the condition where at least three colour criteria are required. While computationally efficient, these methods tend to have conservative classification boundaries, missing potentially interesting objects, that generally lay close to the boundary conditions. This result aligns with \citet{2018A&A...620A.193V}, where a sample of [NeV] emitters was selected, and traditional AGN selection techniques showed low efficiency in their selection.

\section{Physical property estimation}
\label{section:pp_estimation}

Additional efforts can be made to analyse and derive physical properties for our QSO2 candidate sample, using both spectroscopic and photometric SDSS data. In this section, we explore the use of photoionisation models and SED fitting to further characterise our sample.

\subsection{Physical properties derived from [NeIII]}

Considering that our sample has an active black hole, we can estimate relevant physical properties, such as the black hole mass. Assuming the source is radiating at the Eddington limit, the black hole mass can then be estimated using the following equation:
\begin{equation}
M_E = 8 \times 10^5\, \bigg( \frac{L_{\rm{bol}}}{10^{44} \rm{erg s}^{-1}} \bigg)\, M_{\odot},
\end{equation}
\noindent where $L_{\rm{bol}}$ is the bolometric luminosity. A common approach to estimate $L_{\rm{bol}}$ is to use the [OIII]$\lambda5007$ emission line as a proxy \citep[e.g.][see also references therein]{2018ApJ...859..116K}.

However, our new QSO2 candidates with spectra lie in the so-called redshift desert, a redshift range in which the bright optical and UV lines are outside or near the noisy edges and do not allow for reliable analysis. In the absence of the bright high-ionisation line [OIII]$\lambda5007$, we will instead take advantage of the detected [NeIII]$\lambda3869$ emission line. We used the MAPPINGS 1e \citep{1985A&A...143..334B, 1997A&A...322...73F, 2012A&A...547A..29B} photoionisation model from \citet{2021MNRAS.506.1389M} to compute the [OIII]$\lambda5007/$[NeIII]$\lambda3869$ ratio, and then estimate the [OIII]$\lambda5007$ luminosity from the observed [NeIII]$\lambda3869$ flux. 

The photoionisation model has parameters that are expected to be in the correct range for luminous QSO2s since the \cite{2013MNRAS.435.3306A} sample was used. The photoionisation model parameters from \cite{2021MNRAS.506.1389M}, which are also used in this work, are the following:

\begin{itemize}
    \item ionisation parameter\footnote{$U=\frac{Q}{4\pi r^2 n_H c}$, where $Q$ is the luminosity of ionising photons emitted by the ionising source assuming it is isotropic, $r$ is the distance between the cloud and the ionising source, and $n_H$ is the hydrogen density of the cloud.} = 0.01
    \item spectral index of the ionising continuum\footnote{We assume a power-law spectrum of ionising radiation, $S_{\nu} \propto \nu^{+\alpha}$.}: $\alpha$ = -1.5
    \item high energy cut-off for the ionising continuum of $5\times 10^4 $eV
    \item metallicity\footnote{The gas metallicity is expressed with respect to the solar abundance set of \cite{2010Ap&SS.328..179G}.}: $Z = Z_{\odot}$
    \item hydrogen density\footnote{Column density of neutral hydrogen, following the results from \citet{1990ApJ...365..487M,1999A&A...351...47V}.}: $n_H$ = 100 cm$^{-2}$
\end{itemize}

We obtained an [OIII]$\lambda5007$/[NeIII]$\lambda3869$ flux ratio equal to $11.9$. The flux-luminosity relation to estimate the [OIII]$\lambda5007$ luminosity can thus be defined as
\begin{equation}
L_{\rm{[OIII]}} = 11.9\, F_{\rm{[NeIII]}}\, 4\pi D_{L}^{2},
\end{equation}
where $D_L$ is the luminosity distance and $F_{\rm{[NeIII]}}$ is the [NeIII]$\lambda3869$ flux.

Next, we estimate the bolometric luminosity, $L_{bol}$, using [OIII]$\lambda5007$. This conversion has been extensively performed in the literature \citep[e.g.][]{Heckman_2004, 2009A&A...504...73L, 2011A&A...533A.128S}. Here, we estimate $L_{\rm{bol}}$ using the relation given by \citet{2017MNRAS.468.1433P}:
\begin{equation}
\log(L_{\rm{bol}}) = (0.5617 \pm 0.0978) \log(L_{\rm{[OIII]}}) + (22.186 \pm 4.164).
\label{eq:lum_est}
\end{equation}

Another important physical property of AGNs is the mass accretion rate, which can be calculated using
\begin{equation}
\dot{M} = \frac{L_{\rm{bol}}}{\eta c^2},
\label{eq:acc_rate}
\end{equation}
\noindent where $\eta$ is the accretion efficiency, $L_{\rm{bol}}$ is the bolometric luminosity, and $c$ is the speed of light. We considered an accretion efficiency of $\eta = 0.1$ \citep{2017ApJ...836L...1T}. 

\begin{table}
\begin{center}
\caption{Summary of the characterisation of the QSO2 candidates.}
\begin{tabular}{@{}*{5}{c}@{}}
\hline 
   & \parbox[t]{1.2cm}{\centering $L_{[OIII]}$ \\ (erg s$^{-1}$)}   &  \parbox[t]{1.2cm}{\centering $L_{bol}$ \\ (erg s$^{-1}$)} & \parbox[t]{1.4cm}{\centering $\dot{M}$ \\ ($M_{\odot}$ yr$^{-1}$)}   & \parbox[t]{1cm}{\centering $M_{BH}$ \\ ($M_{\odot}$)} \\
 \hline
Median & 2.913$\times 10^{42}$  & 1.092$\times 10^{46}$ & 2.431 &  8.737$\times 10^{7}$ \\
Min & 6.599$\times 10^{41}$ & 4.742$\times 10^{45}$ & 1.055 & 3.794$\times 10^{7}$ \\
Max &2.240$\times 10^{43}$ & 3.434$\times 10^{46}$ & 7.642 &  2.747$\times 10^{8}$ \\
\hline 
\end{tabular}
\tablefoot{From left to right, [OIII] luminosity, bolometric luminosity, accretion rate and black hole mass at the Eddington limit.}
\label{tab:phy_qso2_cand}
\end{center}
\end{table}

A summary of the results of the characterisation of the QSO2 candidates is shown in Table \ref{tab:phy_qso2_cand} (see Appendix \ref{appendix:pp_estimation} for the full table). For $L_{\rm{bol}}$, typical values for AGN sources range between $10^{42}-10^{46}$ erg s$^{-1}$ \citep{Honig2007}.  Additionally, the bolometric luminosities are similar to the ones estimated for the QSO2 sample with $10^{46.3}-10^{ 46.8}$ erg s$^{-1}$ \citep{2013MNRAS.435.3306A}. The AGN black hole mass range is roughly between $10^{7} - 10^8 M_{\odot}$ \citep{Laor2000}. Finally, following the criterion in \citet{2003AJ....126.2125Z} for the selection of QSO2, L$_{\rm{OIII}}$/L$_{\odot}$ \textgreater 3 $\times 10^{8}$ L$_{\odot}$, we obtained 47 sources that fulfil the criterion. The remaining five sources have L$_{\rm{OIII}}$/L$_{\odot}$ between $1.72-2.37 \times 10^{8}$ L$_{\odot}$. Thus, the overall physical properties derived for the QSO2 candidates are consistent with being a luminous AGN.

\subsection{SED fitting with CIGALE}
\label{section:fracAGN}
Due to the low S/N or the lack of an SDSS spectrum, conducting a comprehensive spectroscopic analysis for the entire sample was not possible. Thus, for our entire sample of QSO2 candidates we additionally performed an SED fitting analysis using \texttt{Code Investigating GALaxy Emission} \citep[\texttt{CIGALE}][version \texttt{2022.1}]{2005MNRAS.360.1413B, 2009A&A...507.1793N,2019A&A...622A.103B,2022ApJ...927..192Y}. \texttt{CIGALE} performs the SED fit in a self-consistent framework considering the energy balance between UV/optical absorption and IR emission. From the best-fit model, various parameters can be estimated, including parameters related to the AGN or stellar components of the SED. The models used here are described in \citet{2020MNRAS.491..740Y} and the parameters are presented in Sect. \ref{appendix:SED_parameters}.

To estimate the contribution of the AGN to the SED, we used the \texttt{SKIRTOR} model, which conceptualises the torus as comprised of dusty clumps \citep{2012toru.work..170S, 2016MNRAS.458.2288S}. This model considers that $97\%$ of the mass fraction within the AGN is in the form of high-density clumps, with the remaining $3\%$ evenly distributed, while incorporating an anisotropic disk emission.

For our study, the AGN fraction, frac$_{\rm{AGN}}$, is among the most interesting parameters, as it allows one to understand whether an obscured AGN is likely to be present, validating our AGN classification. 
The frac$_{\rm{AGN}}$ is defined as the AGN contribution to the total IR dust luminosity:

\begin{equation}
    \rm{frac_{AGN} = \frac{L_{dust, AGN}}{L_{dust, AGN} + L_{dust, galaxy}}},
\end{equation}
\noindent where $\rm{L_{dust, AGN}}$ and $\rm{L_{dust, galaxy}}$ are the AGN and galactic dust luminosity, integrated over all wavelengths. We allowed $\rm {frac_{AGN}}$ to vary between zero (no AGN) to 0.99 (AGN dominated).  

\begin{figure}
    \centering
    \resizebox{\hsize}{!}{\includegraphics{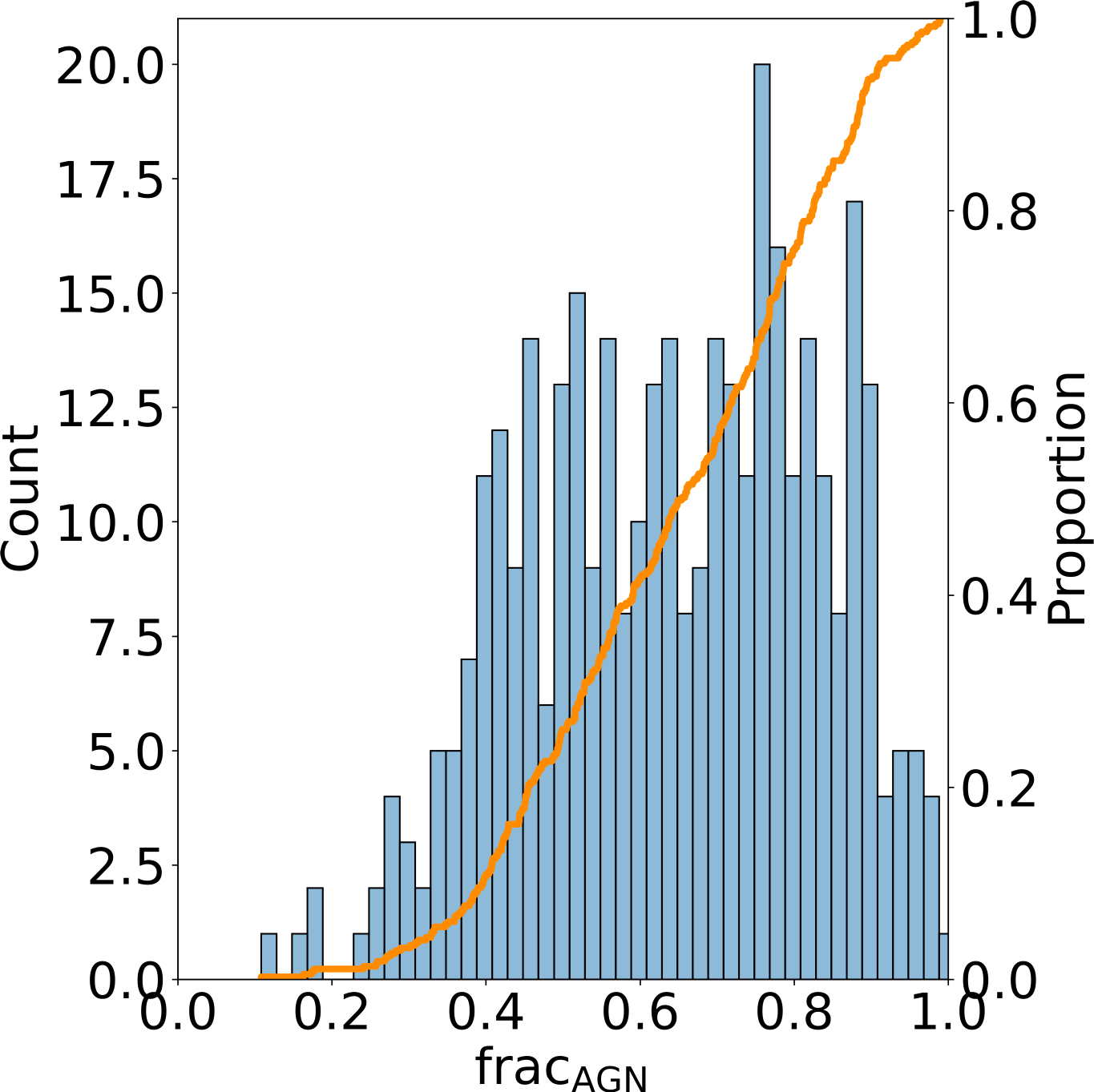}}
    \caption{Histogram (blue) and empirical cumulative distribution function (orange) for the AGN fraction, frac$_{\rm{AGN}}$, derived using \texttt{CIGALE} using the \texttt{SKIRTOR} model. }
    \label{fig:frac_agn}
\end{figure}

Figure \ref{fig:frac_agn} displays the distribution of frac$_{\rm{AGN}}$ obtained using \texttt{CIGALE}. A significant number of QSO2 candidates require high frac$_{\rm{AGN}}$ values, implying that dust re-emission due to a high ionisation source is necessary to explain the observed enhancement in the IR part of the SED. Therefore, including an AGN component to account for the optical-NIR-MIR SED with enhanced dust emission is mandatory to analyse these sources.

A more thorough analysis of the SED fitting results, including a comparison with existing literature, lies beyond the scope of this paper and will be presented in an upcoming publication (Cunha et al., in preparation).

\section{Placing the QSO2 candidates into the galaxy evolution history}
\label{section:context_qso2sample}

Currently, mergers and interactions between galaxies are believed to play crucial roles in the evolutionary history of galaxies and AGNs \citep[e.g.][]{1988ApJ...328L..35S, 2008ApJS..175..356H, 2012ApJ...758L..39T,10.1111/j.1365-2966.2012.20652.x,2018PASJ...70S..37G}. While there is no consensus on the impact of the merger process on the physical properties of obscured AGNs \citep[e.g.][]{2016ApJ...830..156M,2019ApJ...882..141M}, morphological studies have demonstrated its potential as a driving mechanism \citep[e.g.][]{2012ApJ...758L..39T,2019MNRAS.483.1829U,2019ApJ...877...52Z,2020ApJ...904...79M}, in particular for QSO2 \citep[e.g.][]{2009ApJ...702..441G,2011MNRAS.416..262V,2012MNRAS.426..276B,2016MNRAS.457..745W,2023MNRAS.522.5165A,2023MNRAS.522.1736P}.

When two galaxies merge, the first by-product is thought to be a dusty star-forming galaxy which evolves into a dusty AGN. During this transitional phase, the galaxy is heavily obscured by dust, also known as a dust-obscured galaxy \citep[DOG;][]{2008ApJ...677..943D, 2008ApJ...672...94F, 2009ApJ...705..184B, 2009ApJ...700.1190D, 2011ApJ...733...21B,2022ApJ...936..118Y}. Therefore, DOGs are believed to be a transition phase from a gas-rich major merger to an optically thin quasar in a gas-rich major merger scenario \citet{2008ApJ...677..943D}. Due to the dusty environment, DOGs have a very red optical-MIR colour with $i-[W4]_{AB} \geq 7.0$ where the i-band and W4 band are in AB magnitude \citep{2015PASJ...67...86T, 2017ApJ...850..140T, 2018ApJ...857...31T}. Applying this criterion to our new QSO2 candidates, we find that 208 ($\sim 57\%$) of them met the DOG selection criterion. We note that 127 ($\sim 88\%$) of the spectroscopically confirmed QSO2 used for model training (see Sect. \ref{qso_data}) also meet the criterion. This result allowed us to confirm that dust related features are also seen in our candidate sample, validating the presence of high amount of dust, excepted in obscured AGNs.

Extremely red quasars \citep[ERQs; e.g.][]{10.1093/mnras/stv1710, 2017MNRAS.464.3431H, 2020A&A...634A.116V} are of significant relevance in this context. ERQs represent a distinct transition phase population that exhibits both AGN and star-forming activity. \cite{10.1093/mnras/stv1710} identified ERQs using SDSS and WISE photometry, which included various AGN types, including obscured and unobscured QSO, starburst-dominated quasars, and quasars with broad absorption lines. Finally, \cite{2017MNRAS.464.3431H} refined the selection criteria based on broadband photometry and emission line measurements: $i-[W3]_{AB} \geq 4.6$. Interestingly, 38 ($\sim 10\%$) of our new QSO2 candidates meet these criteria (and also meet the DOG criterion). Similarly to the previous analysis, this analysis allowed us to better understand the reliability of this selection, always keeping in mind the different nature of these type of sources from QSO2s.

In summary, we find that most of our QSO candidates are consistent with being dust-obscured host galaxies, as also supported by the results in Sect. \ref{section:fracAGN}. As such, we argue that the sources are likely to be a transitional evolutionary stage, similar to the findings of \citet{2023MNRAS.522..350I} (see discusion in Sect. 4 therein). High-resolution imaging will be crucial to asses this question.

It is tantalising to speculate further on the nature of our QSO2 candidates. The fact that many do not show optical emission lines in their spectra, despite showing infrared colours consistent with the presence of an AGN, might be due to the AGN being so heavily obscured by dust that the ionising UV continuum from the accretion disk is fully absorbed in the nucleus, and does not escape to produce a classical narrow line region. This scenario would represent a relatively early stage in the life of the AGN. 

Furthermore, the dearth of emission lines could mean that our sample contains some analogues of the `line-dark' radio galaxies studied by \cite{2015MNRAS.447.3322H,2016A&A...585A..32H}, where no significant UV continuum or lines were detected \citep[see also][]{2001MNRAS.324....1W}. In a detailed case-study of one such source, \cite{2015MNRAS.447.3322H} concluded that the host galaxy is in an advanced stage of being cleansed of its cold ISM by feedback driven by the AGN, thus representing a relatively late stage in the life of the AGN. One particular source, SDSS J121805.63+583904.6, falls within the definition of a `line-dark' radio galaxy (see Sect. \ref{appendix:Note_individual}). High spatial resolution radio and optical imaging as well as deep optical and IR spectroscopy will be crucial to clarifying the nature and evolutionary stage of our QSO2 candidates. 

\section{Conclusions}
\label{ch6_conc}

We have presented \texttt{AMELIA}, an application of the few-shot machine-transfer-learning based method to identify QSO2s using  optical and infrared photometric data. \texttt{AMELIA} uses a generalised stacking ensemble technique, which acts as an intelligent system of optimisation and model selection, to improve classification performance. We validated the performance of \texttt{AMELIA} for the selection of QSO2s by performing a binary classification of spectroscopically confirmed QSO2s and `normal' galaxies, with an F1-Score $\geq 0.8$ for QSO2 selection. 

We applied the \texttt{AMELIA} pipeline to the problem of identifying QSO2s that were previously misclassified as (non-active) galaxies by the SDSS classification pipeline. We selected 366 new QSO2 candidates, using a class probability of $\geq0.8$. 

From the $138$ candidates that have reliable spectra, we obtained a mean redshift of $z\sim1.1$. Of these sources, $30$ were detected in [NeV]$\lambda3426$ and thus can be considered `spectroscopically confirmed' QSO2s. For an additional $35$ candidates, the [NeIII]$\lambda3869$ line was instead detected, making them `likely' QSO2s. In total, $65$ of our QSO2 candidates show [NeV]$\lambda3426$ or [NeIII]$\lambda3869$, so $\sim47\%$ of the candidates have reliable spectra. 

We cross-matched our candidates with X-ray and radio catalogues and found that ten candidates have X-ray detection and 24 candidates have radio detection. We derived the radio luminosity of these sources and concluded that they show L$_{\rm{150MHz}}$, which is consistent with an AGN source. 

To complement the spectroscopic analysis, we also performed an SED fitting using the \texttt{CIGALE} code. We found that all the QSO2 candidates require an AGN component to account for their SED, corroborating their AGN classification and dusty nature. In fact, more than half of our QSO2 candidates (208) meet the DOG selection criterion, and a significant number (38) also meet the ERQ criterion.

The physical properties of the AGNs were also derived by extrapolation from the [NeIII] $\lambda3869$ flux. We obtained the following median values: L$_{\rm{OIII}} = 2.9\times10^{42}$ erg $s^{-1}$; $L_{\rm{bol}} = 1.1\times10^{46}$  erg $s^{-1}$; $\dot{M} = 2.4$ $M_{\odot}$ yr$^{-1}$; and M$_{\rm BH} = 8.8\times10^7 M_{\odot}$. 

To further characterise our QSO2 candidate sample and confirm our classification, more observations are necessary, which would also allow for more reliable estimation of the host galaxy and AGN properties (e.g. star-formation status, stellar mass, outflow activity status). Multi-wavelength observations (e.g. X-ray, IR, sub-millimetre, and radio) along with high-resolution imaging will play a pivotal role in obtaining a better understanding of the nature of our candidates. Finally, we remark that despite the substantial observational challenges involved with the study of sources in the `redshift desert', QSO2s in this redshift regime provide a crucial bridge between obscured quasar activity at `cosmic noon' ($z\sim2$) and in the low-redshift Universe ($z\leq1$), potentially allowing us to gain insights into the evolution of AGN and galaxy co-evolution across cosmic time.

\begin{acknowledgements}
PACC also thanks Daniel Vaz, Catarina Marques, Christina Th\"{o}rne and José Fernández for the very useful comments and suggestions. PACC acknowledges financial support by Fundação para a Ciência e Tecnologia (FCT) through the grant 2022.11477.BD, and through the research grants UIDB/04434/2020, UIDP/04434/2020 and EXPL/FIS-AST/1085/2021. AH acknowledges supported by Fundação para a Ciência e Tecnologia (FCT) through grants UID/FIS/04434/2019, UIDB/04434/2020, UIDP/04434/2020 and PTDC/FIS-AST/29245/2017, and an FCT-CAPES Transnational Cooperation Project. SGM acknowledges support from Fundação para a Cência e Tecnologia (FCT) through the Fellowships PD/BD/135228/2017 (PhD::SPACE Doctoral Network PD/00040/2012), POCH/FSE (EC) and COVID/BD/152181/2021. SGM was also supported by Fundação para a Ciência e Tecnologia (FCT) through the research grants UIDB/04434/2020 and UIDP/04434/2020, and an FCT-CAPES Transnational Cooperation Project ”Parceria Estratégica em Astrofísica Portugal-Brasil”. R.C. acknowledges support from the Fundação para a Ciência e a Tecnologia (FCT) through the Fellowship PD/BD/150455/2019 (PhD:SPACE Doctoral Network PD/00040/2012) and POCH/FSE (EC) and through research grants PTDC/FIS-AST/29245/2017, EXPL/FIS-AST/1085/2021, UIDB/04434/2020, and UIDP/04434/2020. JB acknowledges financial support from the Fundação para a
Ciência e a Tecnologia (FCT) through national funds PTDC/FIS-AST/4862/2020
and work contract 2020.03379.CEECIND. A.P.A. acknowledges support from the Fundação para a Ciência e a Tecnologia (FCT) through the work Contract No. 2020.03946.CEECIND. AH acknowledges support from NVIDIA through an NVIDIA Academic Hardware Grant Award.

This publication makes use of data products from the Wide-field Infrared Survey Explorer, which is a joint project of the University of California, Los Angeles, and the Jet Propulsion Laboratory/California Institute of Technology, funded by the National Aeronautics and Space Administration. Funding for the Sloan Digital Sky Survey IV has been provided by the Alfred P. Sloan Foundation, the U.S. Department of Energy Office of Science, and the Participating Institutions. SDSS-IV acknowledges support and resources from the Center for High Performance Computing  at the University of Utah. The SDSS website is www.sdss.org. SDSS-IV is managed by the Astrophysical Research Consortium for the Participating Institutions of the SDSS Collaboration including the Brazilian Participation Group, the Carnegie Institution for Science, Carnegie Mellon University, Center for Astrophysics | Harvard \& Smithsonian, the Chilean Participation Group, the French Participation Group, Instituto de Astrof\'isica de Canarias, The Johns Hopkins University, Kavli Institute for the Physics and Mathematics of the 
Universe (IPMU) / University of Tokyo, the Korean Participation Group, Lawrence Berkeley National Laboratory, Leibniz Institut f\"ur Astrophysik Potsdam (AIP),  Max-Planck-Institut f\"ur Astronomie (MPIA Heidelberg), Max-Planck-Institut f\"ur Astrophysik (MPA Garching), Max-Planck-Institut f\"ur Extraterrestrische Physik (MPE), National Astronomical Observatories of China, New Mexico State University, New York University, University of Notre Dame, Observat\'ario Nacional / MCTI, The Ohio State University, Pennsylvania State University, Shanghai Astronomical Observatory, United Kingdom Participation Group, Universidad Nacional Aut\'onoma de M\'exico, University of Arizona, University of Colorado Boulder, University of Oxford, University of Portsmouth, University of Utah, University of Virginia, University of Washington, University of Wisconsin, Vanderbilt University, and Yale University.

LOFAR is the Low Frequency Array designed and constructed by ASTRON. It has observing, data processing, and data storage facilities in several countries, which are owned by various parties (each with their own funding sources), and which are collectively operated by the ILT foundation under a joint scientific policy. The ILT resources have benefited from the following recent major funding sources: CNRS-INSU, Observatoire de Paris and Université d'Orléans, France; BMBF, MIWF-NRW, MPG, Germany; Science Foundation Ireland (SFI), Department of Business, Enterprise and Innovation (DBEI), Ireland; NWO, The Netherlands; The Science and Technology Facilities Council, UK; Ministry of Science and Higher Education, Poland; The Istituto Nazionale di Astrofisica (INAF), Italy. This research made use of the Dutch national e-infrastructure with support of the SURF Cooperative (e-infra 180169) and the LOFAR e-infra group. The Jülich LOFAR Long Term Archive and the German LOFAR network are both coordinated and operated by the Jülich Supercomputing Centre (JSC), and computing resources on the supercomputer JUWELS at JSC were provided by the Gauss Centre for Supercomputing e.V. (grant CHTB00) through the John von Neumann Institute for Computing (NIC). This research made use of the University of Hertfordshire high-performance computing facility and the LOFAR-UK computing facility located at the University of Hertfordshire and supported by STFC [ST/P000096/1], and of the Italian LOFAR IT computing infrastructure supported and operated by INAF, and by the Physics Department of Turin university (under an agreement with Consorzio Interuniversitario per la Fisica Spaziale) at the C3S Supercomputing Centre, Italy.

This research has made use of the VizieR catalogue access tool, CDS,
Strasbourg, France \citep{10.26093/cds/vizier}. The original description 
of the VizieR service was published in \citet{vizier2000}.

In preparation for this work, we used the following codes for Python: Numpy \citet{harris2020array}, Scipy \citep{2020SciPy-NMeth}, Astropy \citep{astropy:2013,astropy:2018,astropy:2022}, Scikit-Learn \citep{scikit-learn}, Pandas \citep{mckinney-proc-scipy-2010}, XGBoost \citep{chen2016xgboost}, CatBoost \citep{prokhorenkova2018catboost}, LightGBM \citep{Ke2017LightGBMAH}, specstack \citep{2019ascl.soft04018T}, matplotlib \citep{Hunter:2007}, seaborn \citep{Waskom2021}.
\end{acknowledgements}

\section*{Data availability}

All data used within this work are publicly available in the SDSS archive. The code is publicly available on GitHub (\url{https://github.com/pedro-acunha/AMELIA}). The complete candidates data set are available from the corresponding author upon reasonable request.

\bibliographystyle{aa} 
\bibliography{46426corr} 

\appendix

\section{The impact of the training data distribution}
\label{section:training_dist}

\begin{figure*}
    \centering
    \includegraphics[width=\linewidth]{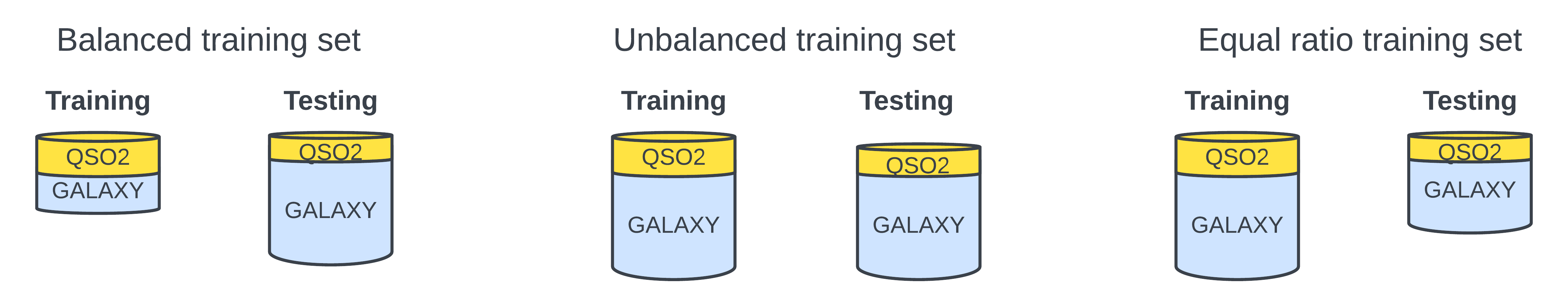}
    \caption{Visual representation of the three different training scenarios. From left to right: {Balanced training set}, \texttt{unbalanced training set}, and \texttt{equal ratio training set}. The yellow colour represents the QSO2 sources, and the blue colour indicates SDSS galaxies.}
    \label{fig:training_sets}
\end{figure*}

We explored two distinct methods for the training-test split and assessed their implications within the semi-supervised one-class approach. We established three different training scenarios for the training set, as illustrated in Fig. \ref{fig:training_sets} for visual reference:

\begin{itemize}
    \item \texttt{balanced training set}: The balanced training set comprises 100 QSO2 sources and 100 galaxy sources. The QSO2 data is distributed in a $7:3$ ratio, with 70$\%$ allocated to the training set and 30$\%$ to the test set. Conversely, the galaxy data follows a $10:90$ ratio, where 10$\%$ is included in the training set, and 90$\%$ is reserved for the testing set.
    \item \texttt{unbalanced training set}: In the unbalanced training set, there are 100 QSO2 sources and 500 galaxy sources. The QSO2 data maintains a $7:3$ ratio, while the galaxy data adopts a $1:1$ ratio, signifying an equal $50\%$ distribution between the training and testing sets.
    \item \texttt{equal ratio training set}: The equal ratio training set ensures the equality of class ratios across both training and testing sets, maintaining a constant $7:3$ distribution for both QSO2 and galaxy classes.
\end{itemize}

\begin{table*}
\begin{center}
\caption{Percentage of False Positives and False Negatives for the different strategies of train-test splits and threshold values.}
\begin{tabular}{@{}*{5}{c}@{}}
\hline
 & \texttt{Equal Ratio ($\%$)} & \texttt{Balance with $0.5$ ($\%$)} &  \texttt{Balance with $0.8$ ($\%$)} & \texttt{Unbalance ($\%$)} \\ \hline
False Positive & 2 & 14 & 9 & 3\\
False Negative & 17 & 5 & 9 & 18  \\
\hline
\end{tabular}
\tablefoot{False positive (galaxy) and false negative (QSO2) comparison for the XGBoost algorithm: Equal ratio training set,  balanced training set with threshold $50\%$ and $80\%$, and unbalanced training set. The results are presented in percentage based on the total number of objects in each classification class.}
\label{table:false_training_clf}
\end{center}
\end{table*}

To explore the implications of various configurations in one-class semi-supervised tasks, as detailed in Sect.\ref{section:semi_supervised}, our investigation focusses on the evaluation of two key metrics. Specifically, we concentrate on evaluating the total number of false positives (FP) for the galaxy class (sources labelled as `galaxy' but classified as QSO2) as well as the false negatives (FN) for the QSO2 class (sources labelled as `QSO2' but classified as galaxy). Table \ref{table:false_training_clf}  displays the analysis of four distinct training data distributions utilising the XGBoost algorithm.

\begin{figure}[h!]
    \centering
    \includegraphics[width=0.95\linewidth]{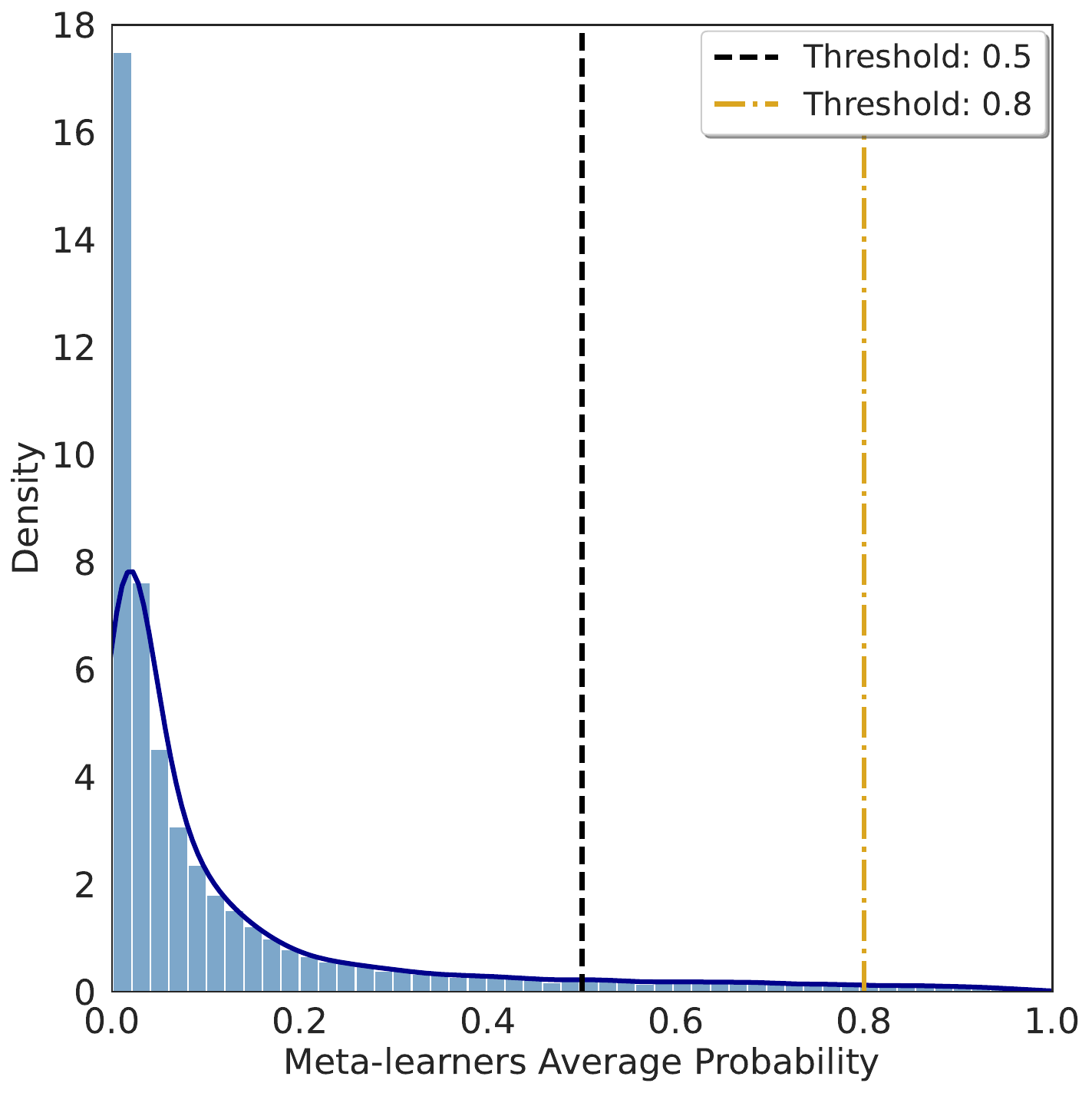}
    \caption{Histogram with the average probability predictions from the generalised stacking method with meta-learner correction. The black dashed line shows the `default value', 0.5, and the yellow dashed line shows the chosen classification threshold in this study, 0.8.}
    \label{fig:hist_meta_preds}
\end{figure}

The balanced training set exhibits a lower FN percentage, indicating greater efficiency in identifying QSO2 sources. These results hold significant value as they aid in selecting the most suitable training distribution for a given task. For instance, the equal ratio training set can be used for selecting galaxies, while the balanced training set demonstrates more promising results for selecting QSO2 sources. This result is not surprising, as it is well-established in the machine learning community \citep[e.g.][and references herein]{lemaavztre2017imbalanced}.

The primary goal of our machine learning pipeline is to identify QSO2 sources within a data set that includes known and unknown classification labels. In the context of dealing with misclassified data in binary classification, false positive (FP) is the most relevant metric for analysis. As the volume of test data increases, the number of FPs also increases. Therefore, we opted for the balanced training set with a threshold $\geq 0.8$, as it achieves a balance between FPs and false negatives (FNs), as seen in Table \ref{table:false_training_clf}. In Fig. \ref{fig:hist_meta_preds}, we show the distribution of the average classification probability from our meta-learners. Since this is a semi-supervised approach, it is not possible to conduct any threshold optimisation analysis, as the ground-truth is not known. Choosing a lower threshold, between 0.5 and 0.8, would substantially increase the number of targets to validate. Our choice, threshold $\geq$ 0.8, enables us to mitigate the bias of our model by balancing the number of misclassifications, due to the increase of the classification threshold, while reducing the number of potential candidates for further validation.

\section{Feature importance analysis}

One of the primary advantages of employing a decision-tree-based algorithm lies in its capability to identify crucial features. These models can quantify the utility of individual features (or observables) in selecting QSO2 sources.

\begin{table}
    \centering
    \caption{Top five features with the highest feature importance for the\texttt{RandomForest}, \texttt{XGBoost}, \texttt{CatBoost}, and \texttt{LightGBM} models.}
    \begin{tabular}{c|c}
    \hline
        Model & Features \\
    \hline
        \texttt{RandomForest} & g-z, g, r-W1, g-W1, g-W2 \\
        
        \texttt{XGBoost} & W4, u-g, u-r, g, r-z \\
        
        \texttt{CatBoost} & g, u-g, W4, g-i, r \\
        
        \texttt{LightGBM} & u-g, g, W4, r-z, r-W4\\
     \hline  
    \end{tabular}
    \tablefoot{The features' order is from the highest feature importance metric to the lowest. To improve the readability, the feature names were simplified.}
    \label{tab:feat_import_top}
\end{table}

We provide a quick analysis of the feature importance provided by \texttt{RandomForest}, \texttt{LightGBM}, \texttt{CatBoost}, and \texttt{XGBoost} models. The majority of the methodologies use different techniques to compute the feature importances. While their examination can inform about the relevant features, their interpretation should be taken with a grain of salt. In the \texttt{RandomForest} algorithm, the feature importance is impurity-based, meaning that it is calculated based on the mean and standard deviation of the impurity measurement from top to bottom of the tree. \texttt{LightGBM} can compute feature importance based on \texttt{split} and \texttt{gain}. In this work, we used the \texttt{split} method which is based on the number of times the feature is used within the model to compute the final prediction. For the \texttt{CatBoost} algorithm, we used the \texttt{PredictionValuesChange} method, which computes the average change in the predicted values if the feature value changes. Finally, for \texttt{XGBoost} the \texttt{weight} importance was used, which computes the number of times a feature appears in a tree, similar to the method used in \texttt{LightGBM}. 

The top five features with the highest importance of the features are summarised in Table \ref{tab:feat_import_top}, ordered by decreasing importance. Some features, such as u-g, g, and W4, appear as some of the most important features in different models. Even though some features appear as less important, it does not mean they do not make a difference in the final classification. We inferred that it is the balance between optical-infrared information that allows the machine learning models to distinguish between the two classes successfully. Some studies explore the relationship between feature importance and final predictions in more depth \citep[e.g.][]{2022A&A...666A.176A,2022MNRAS.512.1710H}. We did not perform such an analysis due to the limited size of our training set.

In \citet{2023A&A...671A..99E}, the authors conducted A/B tests (randomised experiments where one or more feature is removed and the performance model re-evaluated) to understand the impact of features in the classification models, while doing cross-validation and randomly changing random seeds. They have shown that removing multiple features (e.g. broad-band colours, magnitudes) impacts negatively the F1-scores metrics when classifying quiescent galaxies, in the context of Euclid. We performed similar A/B tests for the selection of QSO2 and found that removing features independently of their importance in individual learners has a negative impact.

\section{Performance prediction functions to estimate statistical metrics}

Since we do not have access to ground-truth labels for our test set, it is interesting to quantify the performance of the model based on the model output. In \citet{2022MNRAS.517L.116H}, the authors used a Confidence-Based Performance Estimation (CBPE)\footnote{\url{https://nannyml.readthedocs.io/en/stable/how_it_works/performance_estimation.html}} method from the \texttt{NANNYML}\footnote{\url{https://github.com/NannyML/nannyml}} package to predict the impact of population shifts in unlabelled data for the selection of quasars using decision-tree ensemble models. 

Therein, performance prediction functions allow the estimation of the following metrics for our semi-supervised approach: F1-score, precision, accuracy, recall and specificity. For the performance prediction metrics, we obtain A$=0.90$, P$=0.92$, R$=0.26$, F1-Score$=0.40$ and specificity of $0.997$. Low recall can be explained with the high-threshold setup.

\section{QSO2 candidate with wrong spec-z estimation: SDSS J013056.89-022638.1}
\label{appendix:source_wrongz}

\begin{figure*}
\centering
\includegraphics[width=1\linewidth]{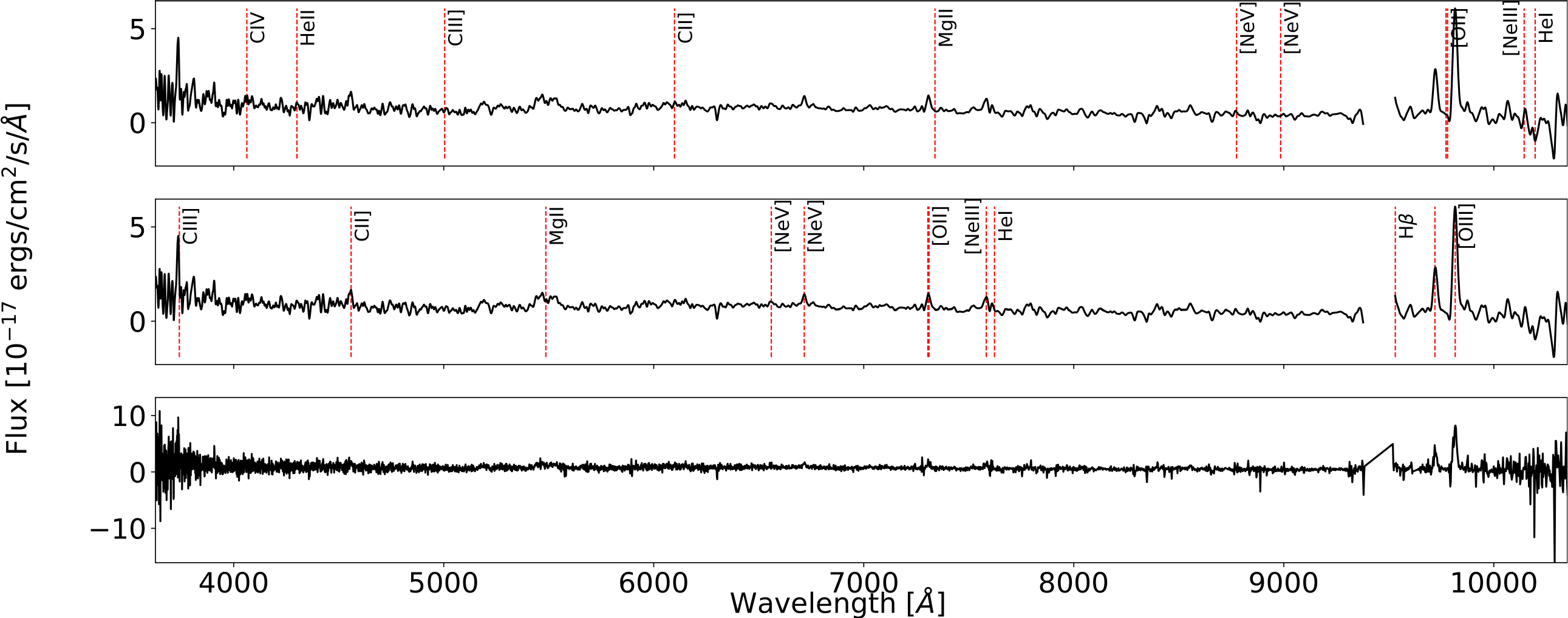}
\caption{Spectrum of SDSSJ013154.44 + 224437.9 from the BOSS spectrograph. In the top plot, the red vertical lines indicate the emission lines for z$\approx 1.633$, computed by \texttt{idlspec2d} from SDSS. In the middle plot, the red vertical lines indicate the emission lines for z$\approx 0.960$, computed by the authors. In the bottom plot, the spectrum without any smoothing is shown.}
\label{fig:spectrum_wrong_z}
\end{figure*}

In our QSO2 candidate sample, we noticed that the spectrum for the source SDSS J013056.89-022638.1 showed emission lines that did not match the estimated SDSS spectroscopic redshift. In particular, [OII]$\lambda\lambda3727,3730$ and [OIII]$\lambda\lambda4959,5008$ were misidentified. By computing the redshift using the [OIII]$\lambda\lambda4959,5008$ doublet, we obtain a spectroscopic redshift z$\approx 0.960$, quite different from z$\approx 1.633$ (computed with \texttt{idlspec2d}). In Fig.\ref{fig:spectrum_wrong_z}, the top plot shows where the emission lines would be placed with z$\approx 1.633$, and the middle plot shows the wavelength for the emission lines with the estimated redshift z$\approx 0.960$. One can easily identify the following emission lines: MgII$\lambda2799$, [NeV]$\lambda\lambda3347,3426$, [OII]$\lambda\lambda3727,3730$, [NeIII]$\lambda3869$, HeI$\lambda3889$, and [OIII]$\lambda\lambda4959,5008$ doublets.
\section{Notes on individual object: SDSS J121805.63+583904.6}
\label{appendix:Note_individual}
Previously identified in the Millions of Optical Radio/X-ray Associations (MORX) v2 catalogue \citet{2024OJAp....7E...6F}, this faint optical source (u: 22.594; g: 23.394; r: 22.368; i: 20.782; z: 19.634) exhibit double radio lobes in the image mosaics from LoTSS DR2. The SDSS spectrum shows no emission lines, being an unclear source, assuming our spectroscopic classification (see Sect.\ref{spec_analysis_QSO2_cand}). Also, this source is consistent with a featureless spectrum, making it a good candidate for a `line-dark' radio galaxy \citep{2015MNRAS.447.3322H,2016A&A...585A..32H}. Even though, it is not possible to confirm the redshift, from the lack of strong AGN emission lines one can extrapolate that this source could be within the `redshift desert'.   

\begin{figure}\centering
\subfloat[SDSS optical image]{\label{a}\includegraphics[width=.45\linewidth]{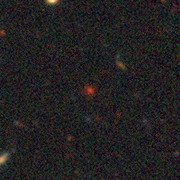}}\hfill
\subfloat[LoTSS radio image]{\label{b}\includegraphics[width=.45\linewidth]{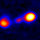}}\par 
\subfloat[SDSS optical spectrum]{\label{c}\includegraphics[width=1\linewidth]{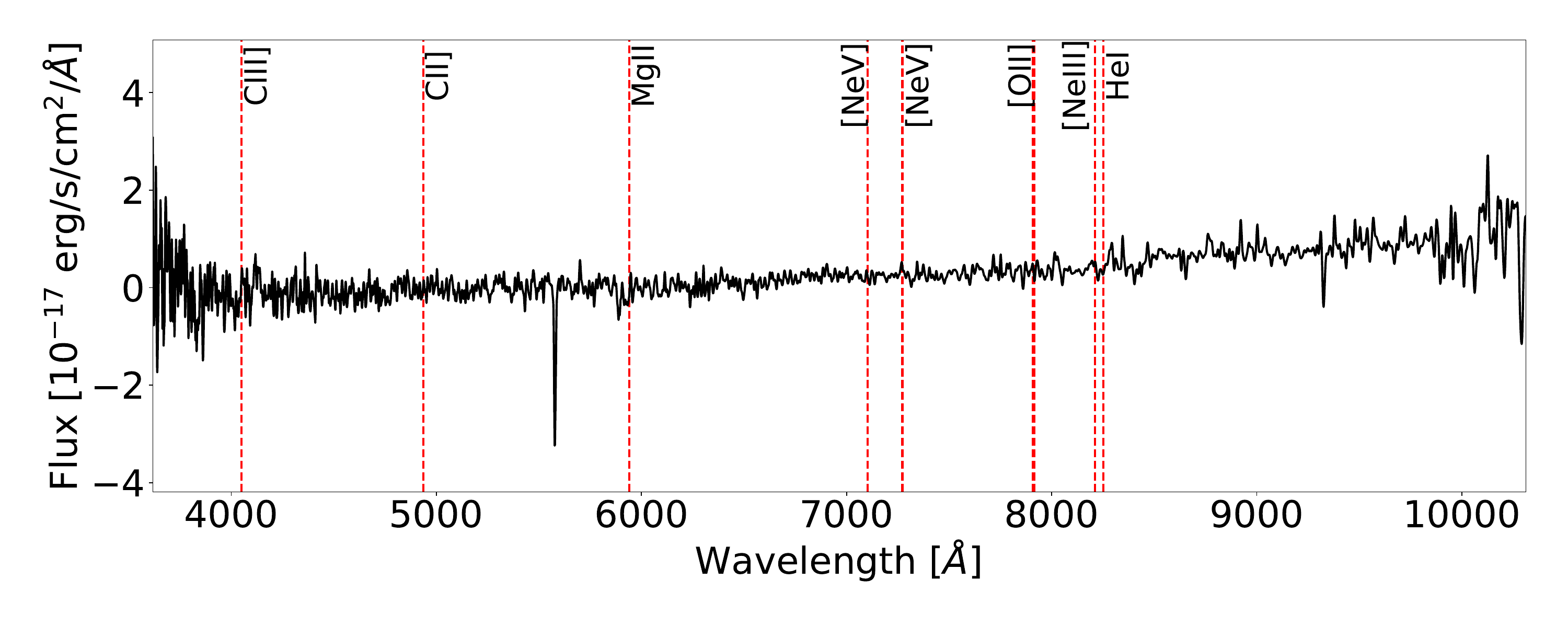}}
\caption{Optical and radio imaging and spectrum for the SDSS J121805.63+583904.6. From left to right, top to bottom: (a) grz cutout image from SDSS with 1$\arcmin$ size; (b)  LoTSS DR2 cutout image with 1$\arcmin$ size; (c) SDSS optical spectrum from BOSS spectrograph, assuming z$=1.122$ .}
\label{fig}
\end{figure}
\section{CIGALE SED fitting parameters}
\label{appendix:SED_parameters}
\begin{table*}
\begin{center}
\caption{Summary of the CIGALE SED fitting parameters used in this work.}
\begin{tabular}{@{}*{3}{c}@{}}
\hline 
Module & Parameter &  Values  \\
\hline 
 Star formation history: & $e$-folding time $\tau_{\rm{main}}$ (Myr) & 5-20 (step 5), 30, 50, 100-300 (step 100),\\
 & &  500.0, 1000.0, 3000.0, 5000.0 \\
 \texttt{sfhdelayed} & Stellar Age, t$_{\rm{main}}$ (Myr)  & 31, 35, 40, 50, 75, 100-500 (step 100), \\
 & & 1000, 3000, 4000, 5000, 5500\\
 & $e$-folding time $\tau_{\rm{burst}}$ (Myr) & 5, 10, 15, 20, 25, 50 \\
 &  Stellar Age, t$_{\rm{burst}}$ (Myr)& 5, 10, 15, 20, 25, 30\\
 & Age of the last burst (Myr) & 0.0, 0.1, 0.2 \\
 \hline\vspace{2mm}
 Galactic dust emission: & $\alpha$ slope & 0.5, 1, 1.5, 2.0, 2.5, 3, 3.5\\
Dale et al. (2014) &  & \\
  \hline 
 AGN: \texttt{SKIRTOR}  & Average edge-on optical depth at 9.7 $\mu$m & 3, 7, 11 \\
 & AGN contribution to IR luminosity, frac$_{\rm{AGN}}$ & 0.0-0.9 (step 0.1), 0.99\\
  Stalevski et al. (2016) & inclination (i.e. viewing angle) & 40, 60, 80\\
  & Intrinsic disk type & Schartmann (2005) spectrum \\
  & Polar extinction E(B-V) &  0, 0.05, 0.1, 0.2, 0.3 \\
\hline 
\end{tabular}
\label{tab:SED_fitting_params}
\end{center}
\end{table*}
\section{Estimation of physical properties}
\label{appendix:pp_estimation} 
\begin{table*}
\caption{Physical properties derived for the [NeIII] emitters using \citet{2021MNRAS.506.1389M} photoionisation models.}
\begin{tabular}{@{}*{7}{c}@{}}
\hline
Plate-MJD-Fiber & z & \parbox[t]{2cm}{\centering L$_{\rm{[OIII]}}$ \\ ( erg $s^{-1}$)}  & L$_{\rm{[OIII]}}$/L$_{\odot}$ & \parbox[t]{2cm}{\centering L$_{\rm{bol}}$ \\ ( erg $s^{-1}$)}  &  \parbox[t]{3cm}{\centering $\dot{M}$ \\ ( M$_{\odot}$ yr$^{-1}$)}   & \parbox[t]{2cm}{\centering M$_{BH}$\\ ( M$_{\odot}$)}  \\
\hline
10447-58143-90  & 1.133 & 1.676E+42 &4.378E+08 & 8.005E+45 & 1.781 & 6.404E+07  \\
10454-58135-808 & 1.109 & 9.084E+41 & 2.373E+08& 5.675E+45 & 1.263 & 4.540E+07   \\
10456-58136-680 & 1.019 & 7.033E+41 & 1.837E+08& 4.915E+45 & 1.094 & 3.932E+07   \\
10468-58131-630 & 1.215 & 2.936E+42 & 7.670E+08& 1.097E+46 & 2.441 & 8.775E+07   \\
10469-58133-915 & 1.102 & 2.378E+42 & 6.213E+08 & 9.744E+45 & 2.168 & 7.795E+07   \\
10739-58255-860 & 1.562 & 4.064E+42 & 1.062E+09& 1.317E+46 & 2.93  & 1.053E+08   \\
11650-58508-299 & 1.141 & 5.245E+42 & 1.370E+09& 1.519E+46 & 3.381 & 1.216E+08   \\
6932-56397-443  & 1.061 & 1.067E+42 & 2.787E+08& 6.211E+45 & 1.382 & 4.969E+07   \\
7239-56693-902  & 1.236 & 1.437E+42 & 3.756E+08 & 7.344E+45 & 1.634 & 5.875E+07   \\
7242-56628-268  & 1.183 & 6.326E+42 & 1.653E+09 & 1.688E+46 & 3.757 & 1.350E+08   \\
7375-56981-607  & 1.057 & 5.606E+42 & 1.465E+09& 1.577E+46 & 3.51  & 1.262E+08 \\
7574-56945-776  & 1.027 & 3.551E+42 & 9.277E+08& 1.221E+46 & 2.716 & 9.764E+07   \\
7839-56900-431  & 1.006 & 1.936E+42 & 5.057E+08& 8.681E+45 & 1.932 & 6.945E+07   \\
8184-57426-269  & 1.115 & 1.959E+42 & 5.117E+08& 8.738E+45 & 1.944 & 6.990E+07   \\
8227-57427-716  & 1.364 & 1.572E+43 & 4.107E+09& 2.815E+46 & 6.264 & 2.252E+08   \\
8409-57867-708  & 1.084 & 3.721E+42 & 9.720E+08& 1.253E+46 & 2.788 & 1.002E+08   \\
8543-57542-675  & 1.012 & 2.891E+42 & 7.552E+08& 1.087E+46 & 2.42  & 8.698E+07 \\
8769-57338-24   & 1.208 & 2.879E+42 & 7.521E+08& 1.085E+46 & 2.414 & 8.679E+07 \\
8769-57338-810  & 1.073 & 1.919E+42 & 5.012E+08& 8.637E+45 & 1.922 & 6.909E+07   \\
8792-57364-92   & 1.021 & 2.478E+42 & 6.473E+08& 9.971E+45 & 2.219 & 7.977E+07  \\
8954-57453-648  & 1.328 & 3.463E+42 & 9.047E+08& 1.203E+46 & 2.678 & 9.627E+07   \\
8958-57514-103  & 1.157 & 6.697E+42 & 1.750E+09& 1.743E+46 & 3.879 & 1.394E+08   \\
9198-57713-103  & 1.166 & 4.802E+42 & 1.254E+09& 1.446E+46 & 3.218 & 1.157E+08   \\
9199-57684-394  & 1.002 & 1.463E+42 & 3.823E+08& 7.418E+45 & 1.651 & 5.935E+07   \\
9204-57712-580  & 1.055 & 1.495E+42 & 3.906E+08& 7.508E+45 & 1.671 & 6.006E+07   \\
9233-58035-981  & 1.146 & 3.766E+42 & 9.839E+08& 1.262E+46 & 2.807 & 1.009E+08   \\
9312-57784-764  & 1.157 & 2.843E+42 & 7.426E+08& 1.077E+46 & 2.397 & 8.617E+07   \\
7338-57874-237  & 1.002 & 1.608E+43 & 4.200E+09& 2.850E+46 & 6.343 & 2.280E+08   \\
7635-56979-874  & 0.960 & 2.240E+43 & 5.851E+09& 3.434E+46 & 7.642 & 2.747E+08   \\
10474-58143-153 & 1.044 & 6.824E+41 & 1.783E+08& 4.833E+45 & 1.075 & 3.866E+07   \\
9315-57713-481  & 1.091 & 1.377E+42 & 3.598E+08& 7.170E+45 & 1.596 & 5.736E+07   \\
9320-58069-26   & 1.102 & 3.823E+42 & 9.988E+08& 1.272E+46 & 2.831 & 1.018E+08 \\
9321-58069-779  & 1.004 & 3.671E+42 & 9.590E+08& 1.243E+46 & 2.767 & 9.948E+07  \\
9323-58047-417  & 1.095 & 2.151E+42 & 5.618E+08& 9.209E+45 & 2.049 & 7.367E+07   \\
9329-57688-155  & 1.053 & 7.342E+42 & 1.918E+09& 1.835E+46 & 4.084 & 1.468E+08   \\
9331-57711-141  & 1.183 & 1.164E+42 & 3.040E+08& 6.522E+45 & 1.451 & 5.218E+07   \\
9335-57713-514  & 1.01  & 2.670E+42 & 6.975E+08& 1.040E+46 & 2.314 & 8.319E+07 \\
9335-57713-675  & 1.023 & 3.339E+42 & 8.721E+08& 1.179E+46 & 2.623 & 9.431E+07   \\
9343-57715-575  & 1.02  & 3.469E+42 & 9.061E+08& 1.205E+46 & 2.68  & 9.636E+07 \\
9345-57713-444  & 1.128 & 2.225E+42 & 5.812E+08& 9.386E+45 & 2.089 & 7.509E+07   \\
9373-58096-257  & 1.25  & 4.309E+42 & 1.126E+09& 1.361E+46 & 3.028 & 1.089E+08   \\
9379-58081-222  & 1.41  & 7.954E+42 & 2.078E+09& 1.920E+46 & 4.272 & 1.536E+08   \\
9384-58080-706  & 1.011 & 8.429E+42 & 2.202E+09& 1.983E+46 & 4.413 & 1.587E+08   \\
9407-58041-946  & 1.101 & 6.599E+41 & 1.724E+08& 4.742E+45 & 1.055 & 3.794E+07   \\
9416-58098-471  & 1.021 & 2.033E+42 & 5.311E+08& 8.923E+45 & 1.986 & 7.138E+07   \\
9553-57801-634  & 1.237 & 2.242E+42 & 5.858E+08& 9.428E+45 & 2.098 & 7.542E+07   \\
9555-57814-860  & 1.145 & 2.382E+42 & 6.222E+08& 9.752E+45 & 2.17  & 7.802E+07   \\
9557-58098-437  & 1.229 & 1.442E+43 & 3.768E+09& 2.682E+46 & 5.968 & 2.146E+08   \\
9594-58135-741  & 1.351 & 4.930E+42 & 1.288E+09& 1.467E+46 & 3.266 & 1.174E+08   \\
9616-58135-297  & 1.037 & 4.382E+42 & 1.145E+09& 1.373E+46 & 3.056 & 1.099E+08   \\
11078-58437-179 & 1.055 & 1.958E+42 & 5.115E+08& 8.736E+45 & 1.944 & 6.989E+07  \\
\hline
\end{tabular}
\label{table:qso2_phy}
\end{table*}

\end{document}